%% file: main.tex
\documentclass[sigconf]{acmart} 

\setcopyright{acmlicensed}
\copyrightyear{2024}
\acmYear{2024}

\renewcommand\footnotetextcopyrightpermission[1]{}


\input{macro}
\begin{document}

\title{Reducing False Positives in Static Bug Detection with LLMs: \texorpdfstring{\\}{ }An Empirical Study in Industry}

\newcommand\corrauthorfootnote[1]{%
  \begingroup
  \renewcommand\thefootnote{}\footnote{\textsuperscript{*}#1}%
  \addtocounter{footnote}{-1}%
  \endgroup
}

\author{
Xueying Du\textsuperscript{1}, 
Jiayi Feng\textsuperscript{1}, 
Yi Zou\textsuperscript{1}, 
Wei Xu\textsuperscript{2}, 
Jie Ma\textsuperscript{2}, 
Wei Zhang\textsuperscript{2}, 
Sisi Liu\textsuperscript{2}, 
Xin Peng\textsuperscript{1}, 
Yiling Lou\textsuperscript{1} 
}

\affiliation{%
    \institution{\textsuperscript{1}Fudan University}\country{China}  
}

\affiliation{%
    \institution{\textsuperscript{2}Tencent}\country{China}  
}


\authornote{Corresponding author}

\authornote{Authors’ Contact Information: Xueying Du, 21210240012@m.fudan.edu.cn; Jiayi Feng, 23210240148@m.fudan.edu.cn; Yi Zou, zouy25@m.fudan.edu.cn; Wei Xu, davidxu@tencent.com; Jie Ma, jima@tencent.com; Wei Zhang, marserzhang@tencent.com; Sisi Liu, rainnyliu@tencent.com; Xin Peng, pengxin@fudan.edu.cn; Yiling Lou, yilinglou@fudan.edu.cn}

\begin{abstract}
Static analysis tools (SATs) are widely adopted in both academia and industry for improving software quality, yet their practical use is often hindered by high false positive rates, especially in large-scale enterprise systems. These false alarms demand substantial manual inspection, creating severe inefficiencies in industrial code review. While recent work has demonstrated the potential of large language models (LLMs) for false alarm reduction on open-source benchmarks, their effectiveness in real-world enterprise settings remains unclear. To bridge this gap, we conduct the first comprehensive empirical study of diverse LLM-based false alarm reduction techniques in an industrial context at Tencent, one of the largest IT companies in China. Using data from Tencent’s enterprise-customized SAT on its large-scale Advertising and Marketing Services software, we construct a dataset of 433 alarms (328 false positives, 105 true positives) covering three common bug types. Through interviewing developers and analyzing the data, our results highlight the prevalence of false positives, which wastes substantial manual effort (e.g., 10 - 20 minutes of manual inspection per alarm).
Meanwhile, our results show the huge potential of LLMs for reducing false alarms in industrial settings (e.g., hybrid techniques of LLM and static analysis eliminate 94–98\% of false positives with high recall). Furthermore, LLM-based techniques are cost-effective, with per-alarm costs as low as 2.1–109.5 seconds and \$0.0011–\$0.12, representing orders-of-magnitude savings compared to manual review. Finally, our case analysis further identifies key limitations of LLM-based false alarm reduction in industrial settings.  
\end{abstract}


\maketitle

\input{section/intro}
\input{section/related}
\input{section/approach}

\input{section/evaluation}
\input{section/discussion}

\input{section/threats}

\input{section/conclusion}

\bibliographystyle{ACM-Reference-Format}

\bibliography{ref}

\end{document}

%% file: macro.tex
\usepackage{hyperref}
\usepackage{adjustbox}
        
\usepackage{float}
\usepackage{graphicx}
\usepackage{textcomp}
\usepackage{xcolor}
\usepackage{multirow}
\usepackage{mdframed}
\usepackage{balance}
\usepackage{booktabs}
\usepackage{tcolorbox}
\usepackage{makecell}
\usepackage{threeparttable}
\usepackage{colortbl}
\usepackage{subfigure}
\usepackage{hhline}
\usepackage{pifont}
\usepackage{listings}
\usepackage{enumitem}

\usepackage{amsmath,amssymb,amsfonts, amsthm}
\usepackage[noend]{algpseudocode}
\usepackage[linesnumbered,ruled,vlined]{algorithm2e}

\newcommand{\parabf}[1]{\noindent\textbf{#1}}

\newcommand{\rev}[1]{\textcolor{black}{\textbf{}#1}}

\definecolor{ggray}{HTML}{eff0f0}
\definecolor{gggray}{HTML}{E8E8E8}
\definecolor{ggggray}{HTML}{BEBEBE}
\definecolor{myblue}{RGB}{255,255,255}
\definecolor{myyellow}{HTML}{FFF2CC}

\usepackage[utf8]{inputenc}
\usepackage[english]{babel}

\definecolor{myrq}{HTML}{E7F1FA}
\definecolor{myblue}{RGB}{255,255,255}
\definecolor{myyellow}{HTML}{FFF2CC}
\definecolor{myfinding}{HTML}{E7F1FA}
\definecolor{myanswer}{HTML}{FDDECE}

\newcounter{finding}

\newcommand{\distance}{7pt}

%% file: section/intro.tex
\section{Introduction}

Static analysis tools (SATs) play a crucial role in ensuring software security, reliability, and maintainability. They scan codebases without extensive test execution~\cite{cai2023place,cai2022peahen,cai2021canary,shi2020conquering,shi2018pinpoint,sui2012static,vassallo2020developers,yan2018spatio}, facilitating early bug explosion during the software development process. 
To date, both academia and industry have proposed many static analysis tools (e.g., Github CodeQL~\cite{codeql}, CppCheck~\cite{cppcheck}, Facebook Infer~\cite{infer}), which are able to detect diverse bug types (e.g., Null-Pointer-Dereference and Use-After-Free) and have been widely adopted in practice~\cite{llm4sa,llm4pfa,guo2023mitigating}. 

However, SATs often suffer from high false positive rates when applied to large-scale software systems in practice. For example, previous studies~\cite{llm4pfa,llm4sa} show that state-of-the-art open-source SATs (e.g., CodeQL and Infer) exhibit over 95\% false alarm rates when detecting Null Pointer Dereference (NPD) bugs in large-scale projects such as the Linux Kernel. This limitation arises from the need to over-approximate program behaviors to guarantee soundness, producing spurious warnings along infeasible paths. The underlying models of SATs typically abstract away complex features like pointer aliasing, external libraries, and conditional compilation, leading to  mismatches with real-world behavior.  Consequently, especially under heightened security requirements,  the trade-off between soundness and precision favors conservative reporting, resulting in warnings that are formally justifiable but practically infeasible, ultimately overwhelming developers with false alarms. 

In practice, bug alarms reported by SATs would be manually inspected by developers to determine whether they correspond to true bugs or false positives. Such manual inspection is  labor-intensive and time-consuming. Our survey within Tencent (i.e., one of the largest IT companies in China) reveals that developers spend on average more than 10 minutes reviewing one single static bug alarm, with many alarms even requiring multiple rounds of inspection.
Moreover, given the large scale of enterprise software codebase, one single scan can produce hundreds of alarms. Therefore, the large volume of false positives leads to significant waste of manual effort, highlighting the necessity of automatic false alarm reduction techniques.

Traditional false alarm reduction techniques rely on symbolic execution~\cite{li2013software,gadelha2019smt} or model checking~\cite{post2008reducing,valdiviezo2014method}, which suffer from scalability and generalization issues when applied to large-scale enterprise software systems.  Learning-based false positive reduction techniques train neural networks on existing false and true positives, which require massive labeled training data for each bug type and thus have limited generalizability across different bug types or for an enterprise where training data is unavailable.
Recently, given the rapid development of Large Language Models (LLMs), some work~\cite{llm4pfa,llm4sa} proposes to reduce false alarms with LLMs, and their experimental results have revealed the potential of LLMs in reducing false alarms reported by open-source SATs on synthetic and open-source projects (e.g., Juliet~\cite{juliet} and Gawk~\cite{gawk}). Nevertheless, \textbf{it still remains unclear how LLM-based false alarm reduction techniques perform in industry, i.e., for real-world proprietary enterprise software}. In particular, 
(1) \textit{proprietary enterprise software} differs significantly from \textit{open-source software}, especially for the code characteristics, development models, and code review practices. For instance, enterprise codebases typically enforce strict internal standards and follow rapid DevOps driven iterations, whereas open-source projects adopt more flexible, community-defined norms and evolve at a slower pace. Moreover, for code review practices, open-source projects benefit from extensive peer reviews by diverse contributors, while proprietary enterprises tend to have more centralized reviews. These differences may impact the types and distributions of bugs in each environment. (2) \textit{Enterprise-customized SATs} also differ a lot from open-source SATs (e.g., CodeQL). In fact, existing open-source SATs often fall short of meeting the comprehensive and customized needs of enterprises, particularly in terms of efficiency, workflow integration, security, and cost. As a result,  enterprises often develop customized SATs tailored to their specific needs and software. The performance differences between these enterprise-developed and open-source tools can significantly influence the effectiveness of false positive reduction. 

To fill the knowledge gap, in this work, we perform an extensive study to empirically investigate how LLMs perform in reducing static false alarms for large-scale industrial software systems at Tencent~\cite{tencent}. Tencent is one of the largest IT companies in China, which covers diverse businesses such as social media, digital entertainment, and cloud computing with billions of active users~\cite{wechat,cloud}. We collect data from the real-world static bug detection pipeline at Tencent, which includes the bug alarms reported by its enterprise-customized SAT (i.e., \rev{BkCheck}) on the large-scale software of its Advertising and Marketing Services (AMS) business line (i.e., over hundreds of components). Based on the historical data, we construct a real-world industry false alarm reduction dataset of 433 bug alarms (i.e., 328 false positives and 105 true positives), which covers the three most frequent bug types, \rev{including Null Pointer Dereference (NPD), Out-of-Bounds (OOB), and Divide-by-Zero (DBZ)} reported by \rev{BkCheck} in Tencent.  Based on our dataset, we empirically study the performance of LLMs in reducing false alarms for industrial software by answering the following research questions.

\textbf{RQ1 (Prevalence of false positives in industry).} We first perform interviews with Tencent developers to investigate the application practice of SATs. SATs serve as a critical gated check-in during the code review process at Tencent, and all the reported bug alarms would be manually reviewed by developers, with false alarms proceeding to a second-round manual validation to avoid missing any potential bugs. False alarms are not only prevalent (i.e., over 76\%), but also require massive manual inspection effort (i.e., on average 10 - 20 minutes per alarm). 

\textbf{RQ2 (Effectiveness of LLMs in reducing false alarms).} We further evaluate a diverse range of different LLM-based false alarm reduction techniques on Tencent datasets, including (1) basic LLMs, (2) different advanced prompting strategies, (3) hybrid techniques of static analysis and LLMs. Moreover, we also compare LLM-based techniques with traditional learning-based techniques on Tencent datasets. \textit{To the best of our knowledge, this is the most comprehensive evaluation on learning-based false alarm reduction of industrial software.} Our experimental results show the strong potential of LLM-based techniques in reducing false positives for industrial software, e.g., LLMPFA, the hybrid techniques of static analysis and LLMs, eliminates 94\%–98\% of false positives across different backbone LLMs while maintaining high recall.

\textbf{RQ3 (Cost analysis).} We measure the time and money costs of LLM-based false alarm reduction techniques when applied on Tencent datasets. In particular, the average time cost per alarm ranges from 2.1 to 109.5 seconds, and the average expense per alarm is 0.0011\$–0.12\$. Such acceptable time and money costs make LLM-based techniques especially valuable for supporting static false positive reduction in industrial practice, substantially reducing manual inspection effort (i.e., 10 - 20 minutes). 

\textbf{RQ4 (Breakdown and case analysis).} We further perform breakdown analysis and find the effectiveness variety of LLM-based false alarm reduction techniques across different bug categories. In particular, we find that most LLM-based techniques perform best on Divide-by-Zero (DBZ) bugs, while their performance is lowest on Null Pointer Dereference (NPD) bugs. 
Moreover, we perform bad case analysis to identify the limitations of LLMs when applied to enterprise-scale bug detection, such as handling bugs of long code contexts, reasoning complex cascaded constraints, and insufficient semantic understanding. These challenges remain future directions for advancing the effectiveness of LLMs-based false alarm reduction in industry. 

To summarize, this paper makes the following contributions:  

\begin{itemize}[itemsep=2pt,topsep=0pt,parsep=0pt, leftmargin=15pt]

\item \textbf{Industrial Investigation}. We interview developers and analyze real-world data within Tencent. Our analysis highlights the prevalence and serious impact from false positives reported by SATs in industrial settings.  

\item \textbf{Extensive Comparison.} We conduct the first empirical study to extensively investigate LLM-based false alarm techniques in \textit{industrial settings}. Our experiments cover a wide range of LLM-based techniques, including traditional learning-based ones, basic LLMs, advanced prompting strategies, and hybrid techniques of LLMs and static analysis. We systematically evaluate both the effectiveness and costs of LLM-based techniques in 433 real-world bug alarms collected from Tencent. While the dataset cannot be released due to confidentiality, all methods and detailed results are available in our replication package~\cite{package}.

\item \textbf{Insights and Implications}. We perform both qualitative and quantitative analysis to show both the promise and limitations of LLM-based false alarm techniques in industry. Moreover, we further discuss practical implications for both practitioners and researchers.
\end{itemize}

The remainder of this paper is organized as follows. Section~\ref{sec:2} presents related work. Section~\ref{sec:3} introduces the dataset and the detailed research settings. Section~\ref{sec:4} presents the results and analysis. Section~\ref{sec:disccusion} discusses our findings and presents implications for both practitioners and researchers. Section~\ref{sec:threats} highlights the threats to the validity of the study. Finally, Section~\ref{sec:conclusion} concludes the paper.

%% file: section/related.tex
\section{Related Work}
\label{sec:2}

\subsection{Enhancing LLMs in Vulnerability Detection}

\subsubsection{LLM-based Vulnerability Detection}
The majority of existing work focuses on prompt engineering~\cite{10.1145/3639476.3639762,DBLP:journals/corr/abs-2312-05275}, such as chain-of-thought~\cite{wei2023chainofthought, zhang2022automatic} and few-shot learning~\cite{brown2020language}, to facilitate more powerful LLM-based vulnerability detection. Additionally, recent work explores fine-tuning approaches~\cite{DBLP:conf/icse/YangGMH24, DBLP:journals/corr/abs-2401-17010, mao2024effectivelydetectingexplainingvulnerabilities} to enhance LLMs in vulnerability detection. Furthermore, recent effort~\cite{du2024vul} explores using historical vulnerability knowledge for retrieval-augmented generation (RAG) to enhance LLMs' vulnerability detection capabilities.
Beyond these, further studies investigated different integration strategies to combine LLMs and static analysis to build more powerful static bug detection techniques~\cite{li2023hitchhikers, sun2023gpt, Li24Enhancing, 10.1145/3653718, li2024llmassistedstaticanalysisdetecting, DBLP:journals/corr/abs-2503-03586, DBLP:conf/icml/Guo0XS025}. 
Li et al.\cite{li2023hitchhikers} propose a framework that integrates UBITect with LLMs to detect Use-Before-Initialization (UBI) bugs in the Linux kernel. Sun et al.~\cite{sun2023gpt} employ LLMs to assist static analysis in identifying logic vulnerabilities in smart contracts. Li et al.~\cite{li2025iris} propose IRIS, which first leverages LLMs to identify the specifications of source/sinks for static analysis tools CodeQL for repository-level vulnerability detection. Wang et al.~\cite{wang2023boosting} propose InferROI, which employs LLMs to infer resource-oriented intentions in code, combined with a two-stage static analysis approach to inspect control-flow paths for potential resource leak detection.

\subsubsection{Empirical Studies}
Many efforts have been dedicated to evaluating LLMs in bug detection~\cite{DBLP:journals/corr/abs-2311-16169,primevul,llmvulsurvey}, covering diverse benchmarks, LLMs, and metrics. However, these benchmarks are primarily derived from synthetic datasets (e.g., Juliet~\cite{juliet}) and open-source projects, whose bug distributions may differ significantly from those encountered in real-world proprietary enterprise codebases. Furthermore, prior work has not systematically evaluated the ability of LLMs to reduce false positives in static bug detection. This study fills this research gap by providing a comprehensive evaluation of LLMs in the context of false positive reduction in enterprise-level static bug detection.

\subsection{False Positive Elimination in Bug Detection}

\subsubsection{Static Analysis-Based False Positive Elimination}
A number of studies have explored the use of more precise static analysis techniques to identify and eliminate false positives from bug reports. These efforts primarily fall into two categories: model checking-based and symbolic execution-based approaches. Model checking-based approaches ~\cite{post2008reducing,valdiviezo2014method} abstract the program into a finite-state model and use formal verification tools to exhaustively explore all possible execution states in order to determine the feasibility of reported paths. Symbolic execution-based approaches~\cite{li2013software,gadelha2019smt} symbolically execute the program to collect path constraints and verify path feasibility using SMT solvers. 
Despite their precision, both categories face significant challenges. Exhaustively exploring all paths leads to path explosion, and their implementations are often tightly coupled with specific static analyzers or require handcrafted rules for different vulnerability types, resulting in poor generalization and restricting their application to large-scale systems.

\subsubsection{Learning-Based False Positive Elimination}
With the rise of deep learning, several approaches have been proposed to train predictive models that distinguish real bugs from false positives based on features extracted from source code or warning reports. 
Lee et al.\cite{lee2019classifying} train a convolutional neural network (CNN) to learn lexical patterns from code segments that trigger alarms. Kharkar et al.\cite{10.1145/3510003.3510153} adopt a Transformer-based model to identify false positive warnings. However, these supervised learning-based approaches require large-scale labeled training data for each bug type, which limits their generalizability across different bug types or in an enterprise where training data is unavailable. 
Recent work combines LLMs and static analysis information to identify false positives during static bug detection. Wen et al.~\cite{llm4sa} propose LLM4SA, which provides LLMs with all the code contexts around the sink (e.g., callers and callees of the sink and including hundreds of lines of code) into one single prompt. Wang et al.~\cite{llmdfa} propose LLMDFA, which leverages few-shot chain-of-thought prompting to guide LLMs to analyze the path feasibility within each function. 
Du et al.~\cite{llm4pfa} propose LLM4PFA, which integrates static analysis–derived path constraints with agent-driven, context-aware reasoning, enabling the model to more accurately assess source-to-sink reachability and thereby eliminate false positives caused by infeasible paths.
\rev{Li et al.~\cite{buglen} propose BUGLENS, a post-refinement framework that guides LLMs through structured reasoning steps to assess security impact and validate constraints from the source code. This work is parallel to ours.}

\subsubsection{Empirical Studies}
\rev{Guo et al.~\cite{guo2023mitigating} provide a comprehensive review of various methods for mitigating false positive static analysis warnings, highlighting the key challenges in this field based on their survey results. Specifically, their study categorizes false positive elimination methods into five distinct approaches: Statistical Probability, Dynamic Program Testing, Machine learning, and Clustering. 
However, their study does not explore the potential of LLMs in mitigating false positives. Moreover, it focuses on false positives from open-source static analyzers, offering limited insight into the effectiveness and challenges of addressing false positives in enterprise settings. 
In addition, several works conduct evaluations and empirical studies of static analysis tools. Li et al.~\cite{study2java} and Lipp et al.~\cite{study3c} compare and evaluate the effectiveness and performance of JAVA and C SASTs, respectively. Ami et al.~\cite{study1fn} conduct in-depth interviews with 20 practitioners to investigate how developers perceive, select, and use SASTs. Li et al.~\cite{study4} investigate why industrial professionals utilize SAST benchmarks, identify barriers to their usage, and explore potential improvements for existing benchmarks. However, these works do not provide a comprehensive study of SAST false positives and how to systematically eliminate them. Our work fills this gap.}

%% file: section/approach.tex
\section{Research Setting}
\label{sec:3}

The goal of our empirical study is to investigate the prevalence of false positives produced by static analysis tools in industrial codebases and to evaluate the effectiveness of existing learning–based approaches in mitigating these false positives. This section outlines the overall research setting. Section~\ref{sec:dataset} details how we collected real-world false positive data from an internal static analyzer within Tencent. Section~\ref{sec:rq} introduces the research questions that guide this study and the specific settings for each research question.

\input{section/benchmark}

\input{section/rq}

%% file: section/benchmark.tex
\subsection{Data Collection}
\label{sec:dataset}

We collected data from the main repositories of Tencent’s Advertising division, encompassing all projects within the Advertising and Marketing Services (AMS) business line. The dataset covers warnings reported between September 2024 and June 2025, including both real bugs and false positives. The overall data analysis and collection pipeline is summarized in Table \ref{table:benchmark}. In particular, the process involves three key steps: \textbf{(1) Automated scanning.} A static analyzer was applied to the codebases to generate warnings. \textbf{(2) Manual review.} Enterprise code reviewers manually examined the warnings to determine whether they represented real bugs or false positives. \textbf{(3) Data collection.} We collected the verified outcomes from the manual review records, followed by a secondary validation to ensure correctness and data integrity.

\input{picture/dataset}

\parabf{Static Analyzer.} 
Mature open-source and commercial static analysis tools often fall short of meeting the comprehensive requirements of enterprises during code review, particularly with respect to efficiency, workflow integration, security, and cost. As a result, Tencent developed its own static analysis tool, BkCheck, to support large-scale static scanning ensuring both high efficiency and security.
\textbf{BkCheck} is a proprietary static analysis tool developed by Tencent’s CodeCC platform. It covers a wide range of detection rules, including null pointer dereference, logic defects, divide-by-zero errors, and memory leaks. Compared to compiler-based analyzers such as Coverity~\cite{Coverity} and CodeQL~\cite{codeql}, BkCheck does not require code compilation, which allows for faster scans and seamless integration into large-scale industrial projects. It has been widely deployed in numerous flagship Tencent products, including WeChat Pay and many popular Tencent games.

\parabf{Manual Review.} During the code review phase, both developers and professional code reviewers manually inspected the warnings reported by static analyzers. The review process consisted of two independent rounds conducted by different reviewers. In the first round, developers examined each warning together with the corresponding code to determine whether it was a false positive or a real bug. For warnings identified as false positives, developers were required to document the reason; for real bugs, the analyzer-reported root cause was retained. In the second round, dedicated code reviewers validated the results of the first round by re-examining the code, warnings, and the initial annotations. The final outcomes from both rounds were consolidated into specialized bug report forms, which recorded the code snippet, bug category, and final review decision. 
\rev{It is worth noting that this review phase is part of Tencent’s routine code review workflow. As such, the resulting manual annotations reflect the professional judgment of experienced enterprise developers and reviewers regarding false positives and real bugs. Consistent with the “security-first” principle, it is highly unlikely that true positives were inadvertently discarded. In cases where reviewers could not reach a definitive conclusion, the warning was marked as unresolved for follow-up verification by more senior reviewers.}

\parabf{Data Collection.} We constructed our dataset from the finalized bug report forms generated during the above review phase. These forms contained 35 categories of bugs, as well as a portion of uncategorized cases. 
\textit{Bug Types.} For this study, we selected the three most frequently occurring and developer-validated categories: Null Pointer Dereference (NPD), Out-of-Bounds (OOB), and Divide-by-Zero (DBZ). These categories accounted for approximately 52\%, 24\%, and 10\% of all recorded warnings, respectively, covering 86\% in total and highlighting their significant impact on enterprise-scale software systems.
\textit{Filtering Criteria.} We systematically extracted all records of these three categories between September 2024 and June 2025.
\rev{Although our double checks revealed no correctness issues in the review forms, certain entries corresponded to deprecated code or contained incomplete contextual information, which could compromise the fairness of the evaluation. To ensure the rigor and reliability of the dataset, we applied the following filtering criteria:
(1) Records associated with deprecated code were removed. (2) Records with incomplete code information or unavailable dependent context were excluded. (3) Cases with unresolved uncertainty regarding false positive status were discarded. (4) Duplicate records were eliminated.}
\textit{Data Statistics.} After filtering, we obtained a total of 328 false positives and 105 real bugs. Table~\ref{table:benchmark} summarizes the detailed statistics of the resulting dataset.

\input{table/benchmark}

%% file: picture/dataset.tex
\begin{figure}[htb]
	\centering
	\includegraphics[width=1\linewidth]{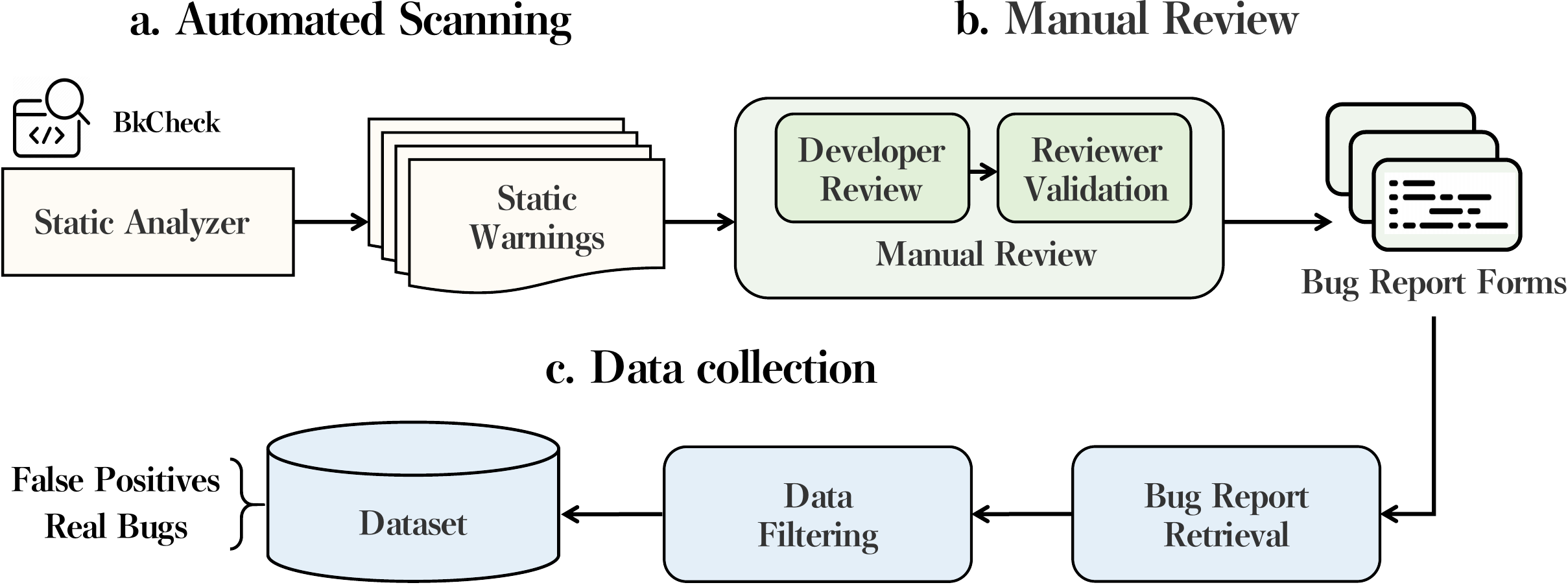}
	\caption{Overview of Data Collection}
	\label{fig:dataset}
 
\end{figure}

%% file: table/benchmark.tex
\begin{table}[htb]
\vspace{-2mm}
    \centering
    \caption{Statistics of Dataset}\label{table:benchmark}
    \footnotesize
    \begin{adjustbox}{width=1\linewidth}
        \begin{tabular}{c|c|c|c}
        \hline
        \textbf{Type} & \textbf{False Alarm Number} & \textbf{Real Bug Number} &
        \textbf{Total}  \\ \hline

        \multirow{1}{*}{NPD}& 107  & 29 & 136 \\\hline

        \multirow{1}{*}{DBZ}& 139  & 19 & 158 \\\hline

        \multirow{1}{*}{OOB}& 82 & 57 & 139     \\\hline

        \multirow{1}{*}{Total}& 328 & 105 & 433    \\\hline
        
        \end{tabular}
        \vspace{-3mm}
    \end{adjustbox}
\end{table}

%% file: section/rq.tex
\subsection{Research Questions}
\label{sec:rq}

Based on the enterprise-level dataset we collected in real-world static bug detection, we addressed the following research questions:

\begin{tcolorbox}[colback=gray!20, colframe=white, width=\linewidth, arc=3mm, boxrule=0.5mm, left=2mm, right=2mm, top=2mm, bottom=2mm, boxsep=0mm]
\textbf{Research Question 1 (Prevalence of false positives in industry):} What is the prevalence of false positives generated by static analysis tools in industrial practice?
\end{tcolorbox}
\vspace{-2mm}

To answer this question, we first perform interviews with Tencent developers to investigate the application practice of SATs, as well as analyzed the collected data. \rev{Specifically, we investigate (1) the industrial use of SATs in Tencent (i.e., how SATs are integrated into the code review workflow), (2) the false positive rates, and (3) the manual effort spent in validating reported false alarms.  }

\rev{\parabf{Interview Protocol.} We designed an online survey questionnaire containing open-ended questions to guide the interviews. The study protocol was approved by the Institutional Review Boards (IRBs) of the authors’ institutions. The complete list of interview questions is provided in our replication package~\cite{package}.}

\rev{\parabf{Interview Recruitment.} We recruited five participants from Tencent’s Advertising and Marketing Services (AMS) business line who were directly involved in software development and code review. The group consisted of two developers and three code reviewers, each with more than five years of professional experience.}

\vspace{-2mm}
\begin{tcolorbox}[colback=gray!20, colframe=white, width=\linewidth, arc=3mm, boxrule=0.5mm, left=2mm, right=2mm, top=2mm, bottom=2mm, boxsep=0mm]
\textbf{Research Question 2 (Effectiveness of LLMs in reducing false alarms):} How effective are a diverse range of LLM-based techniques in reducing false alarms?
\end{tcolorbox}
\vspace{-2mm}

With the rapid advancement of LLMs, they have demonstrated promising performance across a wide range of code-related tasks. Their vulnerability detection capabilities have also been validated on synthetic datasets (e.g., Juliet~\cite{juliet}) and relatively simple open-source function-level datasets (e.g., Devign~\cite{DBLP:conf/nips/ZhouLSD019}). This research question first investigates the practical effectiveness of LLMs in reducing false positives generated by SATs on real-world industrial data. Specifically, we address the following four questions:

\vspace{1mm}
\begin{itemize}[itemsep=2pt,topsep=0pt,parsep=0pt]

\item \textbf{RQ2.a.} What is the effectiveness of directly applying different LLMs?

\item \textbf{RQ2.b.} What is the effectiveness of employing advanced prompting strategies with LLMs?

\item \textbf{RQ2.c.} What is the effectiveness of advanced LLM-based methods combined with static analysis in reducing false positives?

\item \textbf{RQ2.d.} How do LLM-based techniques compare with traditional learning-based techniques in reducing false positives?  

\end{itemize}
\vspace{1mm}

\parabf{Model selection.} To evaluate the impact of different backbone LLMs, we selected four representative state-of-the-art models that are widely adopted in bug detection tasks. These include two closed-source models, GPT-4o~\cite{gpt4o} and Claude-Opus-4~\cite{claude}, as well as two open-source models, Qwen-3-Coder~\cite{qwen} and DeepSeek-R1~\cite{deepseek}.

\parabf{Prompting strategies.} Beyond a basic prompt setting, \rev{we adopt three most widely used and representative prompting strategies following previous research~\cite{prompt4vd1,prompt4vd2,prompt4vd3}.}

\begin{itemize}[itemsep=2pt,topsep=0pt,parsep=0pt]

\item \textbf{Chain-of-Thought (CoT) strategy~\cite{wei2023chainofthought, zhang2022automatic}} explicitly guides LLMs to perform step-by-step reasoning, thereby improving their ability to handle complex logical dependencies when assessing whether a warning is a false positive.

\item \textbf{Few-shot learning strategy~\cite{brown2020language}} provides LLMs with representative examples of both false positives and real bugs. By observing these demonstrations, LLMs learn decision boundaries and make more accurate, expert-like classifications.

\item \textbf{Bug-type augmentation~\cite{zhou2024large}} enriches prompts with structured bug types description information as bug knowledge to LLMs. This strategy supplies LLMs with explicit semantic cues about common bug patterns, enabling more informed reasoning over static analysis warnings.
\end{itemize}

Both few-shot learning and bug-type augmentation can be regarded as forms of domain knowledge enhancement. Bug-type augmentation provides relatively general domain knowledge, while few-shot learning offers concrete representative examples that serve as domain-specific guidance.

\parabf{Hybrid methods between LLM and static analysis.} Recent studies~\cite{llm4pfa,llm4sa} have proposed LLM-based approaches that explicitly integrate static analysis information for false positive reduction. However, these works have only been evaluated on open-source datasets, and their effectiveness in real-world industrial settings remains uncertain. To bridge this gap, we evaluate the following methods on our enterprise dataset:

\begin{itemize}[itemsep=2pt,topsep=0pt,parsep=0pt, leftmargin=15pt]

\item \textbf{LLM4SA~\cite{llm4sa}} leverages LLMs to directly analyze bug warning reports generated by static analysis tools. It supplies the model with extensive code context surrounding the reported sink (e.g., callers, callees, and hundreds of lines of relevant code) and prompts the LLM to determine whether the warning constitutes a false positive.

\item \textbf{LLM4PFA~\cite{llm4pfa}} improves complex inter-procedural path feasibility analysis to reduce false positives reported by static bug analyzers. Specifically, LLM4PFA integrates static analysis–derived path constraints with agent-driven, context-aware reasoning, enabling the model to more accurately assess source-to-sink reachability and thereby eliminate false positives caused by infeasible paths.

\end{itemize}

\rev{Although several other hybrid methods could also be used for false positive reduction, they cannot be directly applied to our target codebase for two reasons: (1) Tool dependency. For example, IRIS~\cite{li2025iris} relies on CodeQL for core taint analysis; however, CodeQL requires a fully compilable repository, which limits its generalizability to our studied large-scale industrial codebase. 
(2) Bug-type constraints. For instance, several bug types in our dataset fall outside the types supported by LLMDFA~\cite{llmdfa}. 
Therefore, we study LLM4SA and LLM4PFA given their easy adoption to our studied industrial codebases. Due to space limitations, we do not include the detailed prompts here. All implementation code for the baselines is available in our replication package~\cite{package}.}

\parabf{Compare with traditional learning-based techniques.}
\rev{To investigate this, we first surveyed review studies evaluating learning-based models for vulnerability detection~\cite{dl4vd1,dl4vd2,dl4vd3} and selected four representative state-of-the-art methods.} These include two graph neural network (GNN)–based approaches and two methods that fine-tune pre-trained language models (PLMs):

GNN-based methods:

\begin{itemize}[itemsep=2pt,topsep=0pt,parsep=0pt, leftmargin=15pt]

\item \textbf{Devign~\cite{DBLP:conf/nips/ZhouLSD019}:} A GNN-based vulnerability detection model that constructs a composite graph representation of source code by integrating multiple program structures, including the abstract syntax tree (AST), control-flow graph (CFG), and data-flow graph (DFG), then applies GNNs to learn semantic features from this graph representation for vulnerability prediction.

\item \textbf{ReGVD~\cite{regvd}:} A GNN-based model combines two GNN variants, ReGGNN and ReGCN, to enhance graph representation learning, thereby improving vulnerability detection accuracy.

\end{itemize}

PLM-based methods:

\begin{itemize}[itemsep=2pt,topsep=0pt,parsep=0pt, leftmargin=15pt]

\item \textbf{LLMAO~\cite{DBLP:conf/icse/YangGMH24}:} A PLM-based fault localization approach that fine-tunes CodeGen on fault localization and vulnerability detection datasets to improve detection precision.

\item \textbf{LineVul~\cite{DBLP:conf/msr/FuT22}:} A PLM-based vulnerability detection model that fine-tunes CodeBERT, offering both function-level and line-level detection capabilities.

\end{itemize}

\vspace{-2mm}
\begin{tcolorbox}[colback=gray!20, colframe=white, width=\linewidth, arc=3mm, boxrule=0.5mm, left=2mm, right=2mm, top=2mm, bottom=2mm, boxsep=0mm]
\textbf{Research Question 3 (Costs Analysis):} What are the costs of using LLMs in false positives reduction?
\end{tcolorbox}
\vspace{-2mm}

This RQ examines the practical costs of deploying LLM-based approaches for false positive reduction in enterprise settings. Although LLMs demonstrate strong potential for mitigating false positives, their adoption in industrial settings must also account for associated costs. We focus on two dimensions of cost:

\begin{itemize}[itemsep=2pt,topsep=0pt,parsep=0pt]

\item \textbf{Response time costs: }The average response time per case for different LLMs and methods.

\item \textbf{Economic costs:} The monetary expense associated with API usage, measured as the average cost per case across models and methods.

\end{itemize}

By analyzing these factors, we aim to provide a comprehensive assessment of the feasibility and sustainability of adopting LLMs for false positive reduction in enterprise environments.

\vspace{-2mm}
\begin{tcolorbox}[colback=gray!20, colframe=white, width=\linewidth, arc=3mm, boxrule=0.5mm, left=2mm, right=2mm, top=2mm, bottom=2mm, boxsep=0mm]
\textbf{Research Question 4 (Breakdown and case analysis):} How do LLMs perform in reducing false positives across different vulnerability categories, and what are their limitations?
\end{tcolorbox}
\vspace{-2mm}

\parabf{Breakdown Analysis.} False positives generated by static analysis tools may differ significantly across vulnerability categories, as each category involves distinct bug patterns and detection challenges. To better understand these variations, we investigate the performance of LLMs in reducing false positives across the three most frequent and impactful categories in our dataset: Null Pointer Dereference (NPD), Divide-by-Zero (DBZ), and Out-of-Bounds (OOB). This research question aims to reveal whether LLMs exhibit consistent effectiveness across categories.

\parabf{Case Analysis.} We perform bad case analysis to better understand the limitations of LLMs in identifying false positives. 
This analysis provides deeper insights into two dimensions:(i) \textit{Error Patterns.} We categorize the representative bad cases where LLMs misdistinguish between false positives and real bugs, and further analyze the conditions under which the LLMs' reasoning capabilities fall short. (ii) \textit{Root Causes.} We further investigate the underlying reasons behind these bad cases, thereby providing a systematic understanding of the inherent limitations of LLMs in static bug detection.

%% file: section/evaluation.tex
\section{Result and Analysis}

\label{sec:4}

\subsection{Metrics and Setup}

For effectiveness evaluation (i.e, RQ2-RQ4), we use four commonly-adopted metrics in bug detection tasks: accuracy, precision, recall, and F1. Additionally, we use two metrics introduced in LLM4PFA~\cite{llm4pfa} to evaluate the effectiveness in identifying false positives generated by static analyzers: False Positive Reduction Precision (FPR\_P) and False Positive Reduction Recall (FPR\_R), which specifically measure the precision and recall for negative samples (i.e., false positives) in the dataset.

$$\text{FPR\_P} = \frac{\text{TN}}{\text{TN} + \text{FN}}   \
\text{FPR\_R} = \frac{\text{TN}}{\text{TN} + \text{FP}}$$

To mitigate potential concerns regarding LLM randomness, we executed each query three times per case and adopted a majority-vote strategy to determine the final outcome. For traditional deep-learning based methods, we directly applied the models trained on the original datasets used in their papers to conduct experiments, since no enterprise-level training dataset is publicly available.

\input{section/evaluation/rq1}

\input{section/evaluation/rq3}
\input{section/evaluation/rq6}
\input{section/evaluation/rq4}

%% file: section/evaluation/rq1.tex
\subsection{RQ1: Prevalence of False Positives}
\label{sec:rq1_result}

\rev{We conducted interviews with five participants and analyzed their responses to the survey questionnaire. Based on this analysis, we derived the following interview results:}

\parabf{Industrial Practice of SATs.} 
As illustrated in Figure~\ref{fig:review}, BkCheck, Tencent’s enterprise-customized SAT, functions as a critical gated check-in by scanning each commit. The reported bug alarms undergo up to two rounds of manual inspection: first, developers manually review all reported alarms, and then those identified as false alarms are subjected to a second-round validation by a different reviewer to minimize the risk of overlooking potential bugs. Overall, SATs play a fundamental role in industrial software quality assurance, with their reported alarms being taken seriously as part of the enterprise code review process.

\input{picture/review}

\parabf{High False Positive Rate.} In our final dataset from Tencent, the proportion of false positives was at 76\% (e.g., 328 false positives and 105 true bugs). In fact, the real-world distribution of false positives could be even higher as a substantial number of false-positive records are excluded from our final dataset due to incomplete or inaccessible code contexts. When accounting for these discarded cases, the false positive rate of static warnings is higher than 90\%.  Such a high false positive rate stems from a design trade-off: enterprise static analysis tools prioritize recall over precision to ensure that no potential bugs are overlooked. Consequently, false positive rates in enterprise settings are even higher than those of many open-source analyzers. 

\parabf{High Manual Effort for Validation.} As shown in Figure~\ref{fig:review}, each reported bug alarm undergoes up to two rounds of manual inspection to minimize the risk of overlooking potential bug. Based on our interviews, the average time required to examine a single alarm was approximately 10 minutes per round. As a result, one single bug alarm would cost 10 to 20 minutes of manual inspection effort.  As hundreds of alarms can be reported by SATs for scanning large-scale enterprise software, significant human effort are wasted for inspecting false alarms in practice. 

\vspace{2mm}
\begin{mdframed}[linecolor=gray,roundcorner=12pt,backgroundcolor=gray!15,linewidth=3pt,innerleftmargin=2pt, leftmargin=0cm,rightmargin=0cm,topline=false,bottomline=false,rightline = false]

\textbf{Answer to RQ1:} SATs play a fundamental role in industrial software quality assurance, with their reported alarms being taken seriously as part of the enterprise code review process. However, false alarms are extremely prevalent (i.e., with 90\% false alarm rate). Meanwhile, substantial manual effort is wasted to inspect reported bug alarms, with one false alarm on average taking 10 to 20 minutes of manual inspection. Consequently, the automatic reduction of false positives in SATs is a critical and urgent need in enterprise code review processes.
\end{mdframed}

%% file: picture/review.tex
\begin{figure}[htb]
	\centering
	\includegraphics[width=1\linewidth]{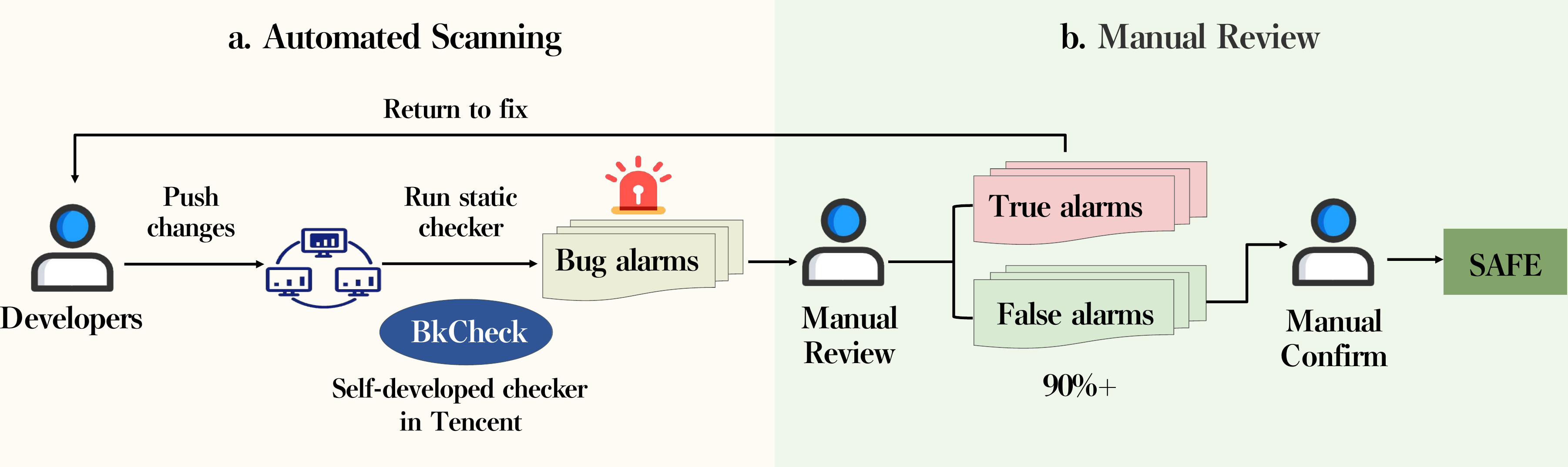}
	\caption{Overview of Code Review in Tencent}
	\label{fig:review}
 
\end{figure}
\vspace{-3mm}

%% file: section/evaluation/rq3.tex
\subsection{RQ2: Effectiveness of LLMs in Reducing False Alarms}
\label{sec:rq3_result}

To investigate the effectiveness of LLMs in reducing false positives, we conduct a comprehensive evaluation of various LLM-based false alarm reduction techniques on our Tencent datasets. Moreover, we compare the performance of LLM-based approaches with that of traditional learning-based methods.

\subsubsection{Effectiveness of various LLM-based false alarm reduction techniques}
Table~\ref{table:rq3} compares the performance of all LLM-based false alarm reduction techniques on our Tencent dataset using various backbone LLMs. 

\input{table/rq3_result}

\parabf{Overall Effectiveness.} LLMs demonstrate strong potential for static false positive reduction, with hybrid methods that combine LLMs and static analysis (i.e., LLM4SA and LLM4PFA) achieving the highest accuracy of 0.93-0.94. These LLM-based approaches also provide a degree of interpretability, making their outputs more available to subsequent human validation.

\parabf{Method Comparison.} Across all evaluated strategies, methods combining LLMs with static analysis (i.e., LLM4SA and LLM4PFA) outperform standalone LLMs with different prompting strategies, which in turn surpass basic prompt–only methods. Among prompting strategies, Chain-of-Thought (CoT) delivers the weakest performance, even underperforming basic prompts in some cases. This may be due to these advanced LLMs have already internalized reasoning capabilities during the training process, and explicitly enforcing CoT prompts can paradoxically degrade performance. In contrast, few-shot enhanced prompting consistently yields the best results among the three prompt strategies, underscoring the value of providing concrete examples to guide LLM reasoning. Both LLM4SA and LLM4PFA not only demonstrate strong effectiveness, with all backbone LLMs achieving accuracy rates above 0.86, but also exhibit greater stability across different backbone LLMs compared to methods relying solely on prompt strategies. These results further validate the potential of integrating LLMs with static analysis for effective false positive reduction.

\parabf{Generalization Across LLMs.} Overall, the four state-of-the-art LLMs exhibit comparable levels of effectiveness, though their relative strengths vary across methods and prompting strategies. For instance, Claude performs competitively with bug-type–enhanced and few-shot–enhanced prompts, but its performance is noticeably weaker with basic prompts. In contrast, Qwen-3-Coder delivers the most consistent performance across all prompting strategies. Importantly, LLM4PFA exhibits strong generalization across all models, achieving accuracy above 0.93 in every case. Given that enterprises typically prioritize recall over precision, LLM4PFA with Claude (recall = 0.88) and DeepSeek-R1 (recall = 0.86) are more suitable for industrial practice.

\subsubsection{Compare with traditional learning-based techniques}
Table~\ref{table:rq2} presents the results of four representative deep learning–based vulnerability detection methods on our Tencent dataset.

\input{table/rq2_result}

\parabf{Overall Effectiveness.} Overall, all evaluated LLM-based methods consistently outperform deep learning approaches. The latter exhibited markedly poor performance, with the highest accuracy reaching only from 0.36 to 0.44 (except for LineVul). Two key factors may account for this limitation. First, deep learning–based methods are highly dependent on training data distribution. These learning-based models trained on existing benchmark datasets perform poorly when directly applied to previously unseen enterprise code, where large-scale, labeled training data is scarce. This lack of sufficient domain-specific training data makes fine-tuning such models for enterprise use particularly challenging.
Second, prior studies~\cite{du2024vul} have shown that learning-based approaches often struggle to capture the true root causes of bugs. Instead, they tend to rely on superficial statistical patterns or secondary code features correlated with buggy functions. As a result, these methods are inherently limited in their ability to generalize to real-world enterprise bug detection.

\parabf{Methods Comparison.} Although Linevul reports a seemingly high accuracy of 0.76, this result stems from its tendency to classify all warnings as false positives. Since our dataset contains a large proportion of false positive cases, this bias inflates its accuracy but undermines practical value. A likely reason is that LineVul was trained on the highly imbalanced BigVul dataset, where most samples are non-vulnerable. By contrast, the other methods all achieve accuracies below 0.44. Among them, ReGVD stands out due to its comparatively high recall of 0.87. However, its false positive recall is only 0.19, meaning that it only successfully eliminates 19\% of false positives. This limited effectiveness in reducing false alarms still imposes substantial manual validation costs, restricting its overall practical benefit.

\parabf{Interpretability.} Traditional learning-based methods inherently suffer from poor interpretability. Their outputs are limited to predicted labels without any explanatory context, which increases the burden of manual validation during enterprise code review. In contrast, LLM-based methods provide natural language explanations alongside predictions. These explanations significantly support human reviewers.

\vspace{2mm}
\begin{mdframed}[linecolor=gray,roundcorner=12pt,backgroundcolor=gray!15,linewidth=3pt,innerleftmargin=2pt, leftmargin=0cm,rightmargin=0cm,topline=false,bottomline=false,rightline = false]
\textbf{Answer to RQ2:} 
LLMs demonstrate strong potential for reducing the false positives reported by SATs in industrial settings. Among all LLM-based approaches, hybrid techniques that integrate LLMs with static analysis (i.e., LLM4PFA) achieve the best effectiveness, successfully eliminating 94–98\% of false positives across different backbone LLMs while maintaining high recall. In contrast, traditional learning-based vulnerability detection methods perform poorly  due to the scarcity of large-scale, high-quality enterprise training data. 
Furthermore, LLM-based methods offer superior interpretability by augmenting label-only predictions with natural language explanations, thereby increasing the practicality and usability of their outputs for subsequent manual validation.
\end{mdframed}

%% file: table/rq3_result.tex
\begin{table*}[htb]
    \centering
    \caption{Effectiveness of LLM-based Methods with different LLMs}\label{table:rq3}
    \footnotesize
    \begin{adjustbox}{width=0.65\linewidth}
        \begin{tabular}{c|c|c|c|c|c|c|c}
        \hline
        \textbf{Technique} & \textbf{Model} & \textbf{Accuracy} &
        \textbf{Precision} & \textbf{Recall} & \textbf{F1} &
        \textbf{FPR\_P} & \textbf{FPR\_R} \\ \hline

        \multirow{4}{*}{Basic Prompt}& GPT-4o & 0.50 & 0.28 & 0.66 & 0.39 & 0.80 & 0.45\\
        & Claude-Opus-4  & 0.40 & 0.26 & 0.83 & 0.40 & 0.83 & 0.26\\
        & Qwen-3-Coder &  0.55 & 0.29 & 0.62 & 0.40 & 0.81 & 0.52 \\
        & DeepSeek-R1  & 0.43  & 0.27 & 0.84 & 0.41 & 0.85 & 0.29 \\ \hline

        \multirow{4}{*}{Cot Enhanced}& GPT-4o & 0.49  & 0.29 & 0.76 & 0.42 & 0.84 & 0.40\\
        & Claude-Opus-4  & 0.42  & 0.29 & \textbf{0.94} & 0.44 & 0.93 & 0.25\\
        & Qwen-3-Coder & 0.48  & 0.29 & 0.79 & 0.42 & 0.85 & 0.38\\
        & DeepSeek-R1  & 0.40  & 0.28 & 0.90 & 0.42 & 0.89 & 0.24 \\ \hline     
        
        \multirow{4}{*}{Few-shot Enhanced}& GPT-4o & 0.50  & 0.26 & 0.64 & 0.37 & 0.79 & 0.43\\
        & Claude-Opus-4  & 0.72  & 0.46 & 0.74 & 0.57 & 0.90 & 0.72\\
        & Qwen-3-Coder &  0.71 & 0.40 & 0.59 & 0.48 & 0.85 & 0.72\\
        & DeepSeek-R1  &  0.69 & 0.43 & 0.83 & 0.56 & 0.92 & 0.65\\ \hline 

        \multirow{4}{*}{Bug Type Enhanced}& GPT-4o &  0.53 & 0.27 & 0.54 & 0.36 & 0.78 & 0.52 \\
        & Claude-Opus-4  & 0.67  & 0.41 & 0.87 & 0.56 & 0.93 & 0.60\\
        & Qwen-3-Coder &  0.65 & 0.37 & 0.64 & 0.47 & 0.85 & 0.65\\
        & DeepSeek-R1  & 0.64  & 0.38 & 0.83 & 0.52 & 0.91 & 0.57\\ \hline 

        \multirow{4}{*}{LLM4SA}& GPT-4o & 0.86  & 0.63 & 0.84 & 0.72 & 0.94 & 0.84\\
        & Claude-Opus-4  & 0.89  & 0.68 & 0.72 & 0.70 & 0.91 & 0.89\\
        & Qwen-3-Coder & 0.92  & 0.85 & 0.65 & 0.74 & 0.90 & 0.96\\
        & DeepSeek-R1  & 0.91  & 0.77 & 0.86 & 0.81 & 0.95 & 0.92\\ \hline 

        \multirow{4}{*}{LLM4PFA}& GPT-4o & 0.93 & \textbf{0.93} & 0.75 & 0.83 & 0.93 & \textbf{0.98}\\
        & Claude-Opus-4  & 0.93  & 0.83 & 0.88 & 0.85 & \textbf{0.96} & 0.94\\
        & Qwen-3-Coder & \textbf{ 0.94} & \textbf{0.93} & 0.79 & \textbf{0.86} & 0.94 & 0.98\\
        & DeepSeek-R1  &  0.93 & 0.84 & 0.86 & 0.85 & 0.95 & 0.95\\ \hline 
        
        \end{tabular}
    \end{adjustbox}
\end{table*}


%% file: table/rq2_result.tex
\begin{table}[htb]
    \centering
    \caption{Effectiveness of traditional learning-based methods}\label{table:rq2}
    \footnotesize
    \begin{adjustbox}{width=1\linewidth}
        \begin{tabular}{c|c|c|c|c|c|c}
        \hline
        \textbf{Tech.} & \textbf{Acc.} &
        \textbf{Pre.} & \textbf{Recall} & \textbf{F1} &
        \textbf{FPR\_P} & \textbf{FPR\_R}\\ \hline

        Devign & 0.44 & 0.35 & 0.61 & 0.44 & 0.60 & 0.34 \\
        \hline 

        ReGVD & 0.36 & 0.26 & 0.87 & 0.39 & 0.82 & 0.19\\
        \hline 

        LLMAO & 0.36 & 0.22 & 0.67 & 0.34 & 0.71 & 0.26 \\
        \hline 

        LineVul & 0.76 & 0.00 & 0.00 & 0.00 & 0.76 & 1.00 \\
        \hline 
           
        \end{tabular}
    \end{adjustbox}
\end{table}

%% file: section/evaluation/rq6.tex
\subsection{RQ3: Costs Analysis}

\label{sec:rq6_result}

In order to provide a comprehensive assessment of the feasibility and sustainability of adopting LLMs for false positive reduction in enterprise environments, Table~\ref{table:rq6} compares the time costs and economic costs of different LLM-based methods using various backbone LLMs.

\input{table/rq6_result}

Overall, LLM-based methods fall well within the acceptable range of both time and economic costs for an enterprise environment. 

\parabf{Time Cost Analysis.} The average processing cost per alarm ranges from 2.1 to 109.5 seconds, depending on the specific method and backbone model. Even the most time-consuming LLM, DeepSeek-R1 (i.e, about 1-2 minutes per warning), remains substantially lower than manual validation, which requires approximately 20 minutes per warning across multiple review rounds.
Moreover, when integrated with DeepSeek, LLM4PFA achieves an accuracy of 0.93 while maintaining a recall of 0.86, making it highly effective for assisting enterprises in false positive reduction, as it can replace at least one round of manual review and save considerable human effort without compromising reliability. 

\parabf{Economic Cost Analysis.} The average expense per alarm is 0.0011\$ – 0.12\$. In addition, two open-source models can be deployed within enterprise infrastructure, eliminating API-related costs. Even the most expensive commercial model (i.e., Claude-Opus-4, 0.12\$ per warning) remains negligible compared to the substantial savings in manual effort, underscoring the usability of LLM-based approaches.

\vspace{2mm}
\begin{mdframed}[linecolor=gray,roundcorner=12pt,backgroundcolor=gray!15,linewidth=3pt,innerleftmargin=2pt, leftmargin=0cm,rightmargin=0cm,topline=false,bottomline=false,rightline = false]
\textbf{Answer to RQ3:} LLM-based false alarm reduction techniques  are cost-effective in industrial settings, incurring both acceptable time and economic costs. On average, the processing time per alarm ranges from 2.1 to 109.5 seconds, while the expense per alarm remains as low as \$0.0011–\$0.12. Compared to the 10–20 minutes typically required for manual inspection, these techniques offer a substantial reduction in developer effort, underscoring their practical value for mitigating false positives in static analysis within industry.

\end{mdframed}

%% file: table/rq6_result.tex
\begin{table}[htb]
    \centering
    \caption{Cost of LLM-based Methods with different LLMs}\label{table:rq6}
    \footnotesize
    \begin{adjustbox}{width=1\linewidth}
        \begin{tabular}{c|c|c|c}
        \hline
        \textbf{Tech.} & \textbf{Model} & \textbf{Ave. time cost (s)} &
        \textbf{Ave. API cost (\$)} \\ \hline

        \multirow{4}{*}{Basic Prompt}& GPT-4o & 5.9 & 0.015 \\
        & Claude-Opus-4  & 12.8 & 0.065\\
        & Qwen-3-Coder  & 2.1 & 0.0013\\
        & DeepSeek-R1  & 109.5 & 0.0046 \\ \hline

        \multirow{4}{*}{Few-shot Prompt}& GPT-4o & 7.9 & 0.020\\
        & Claude-Opus-4  & 14.8 & 0.120\\
        & Qwen-3-Coder & 8.1 & 0.0011\\
        & DeepSeek-R1  & 60.5 & 0.0018 \\ \hline

        \multirow{4}{*}{LLM4SA}& GPT-4o & 8.6 & 0.017\\
        & Claude-Opus-4  & 15.2 & 0.094\\
        & Qwen-3-Coder & 17.2 & 0.0084\\
        & DeepSeek-R1  & 60.2 & 0.0024 \\ \hline

        \multirow{4}{*}{LLM4PFA}& GPT-4o & 12.6 & 0.020\\
        & Claude-Opus-4  & 16.7 & 0.120 \\
        & Qwen-3-Coder & 16.1 & 0.0056\\
        & DeepSeek-R1  & 58.3 & 0.0025 \\ \hline
        
        \end{tabular}
    \end{adjustbox}
\end{table}

%% file: section/evaluation/rq4.tex
\subsection{RQ4: Breakdown and Case Analysis}
\label{sec:rq4_result}

\input{picture/bug_type}

\parabf{Breakdown Analysis. }Figure~\ref{fig:bugtype_4o} illustrates the effectiveness variety of LLM-based false alarm reduction techniques across different bug categories.
Overall, the effectiveness of different LLM-based methods varies across bug categories. Almost all methods achieve their best performance on Divide-by-Zero (DBZ) bugs, largely because the bug patterns in this category are relatively simple and consistent, making them easier for LLMs to identify. In contrast, Null Pointer Dereference (NPD) bugs exhibit more complex and diverse patterns. Moreover, many NPD false positive cases involve intricate conditional constraints, which pose significant challenges for LLMs to reason accurately about program behavior, resulting in comparatively lower performance in this category.

\vspace{2mm}
\begin{mdframed}[linecolor=gray,roundcorner=12pt,backgroundcolor=gray!15,linewidth=3pt,innerleftmargin=2pt, leftmargin=0cm,rightmargin=0cm,topline=false,bottomline=false,rightline = false]
\textbf{Answer to RQ4 (Breakdown Analysis):} The effectiveness of LLM-based methods vary across different bug categories. In Tencent datasets, LLM-based methods achieve the best performance on Divide-by-Zero (DBZ) bugs, while their performance is lowest on Null Pointer Dereference (NPD) bugs. 
\end{mdframed}
\vspace{2mm}

\parabf{Case Analysis.} To better understand the limitations of LLMs in reducing false positives, we conducted a manual analysis of two types of failure cases: (1) cases where different prompt strategies failed to help LLMs distinguish between false positives and real bugs, and (2) cases where all methods failed to classify correctly. Although confidentiality constraints prevent us from presenting the actual code examples, our analysis reveals several key limitations of LLMs in bug detection.

\textit{(i) Limited capacity for long-context reasoning.} We observed that LLMs are significantly less effective at mitigating false positives in cases involving long code contexts. For cases where all methods failed, the average length of the function containing the warning was 95.6 lines longer than the overall dataset average, and 67.7 lines longer than cases only correctly classified by LLM4PFA but missed by other LLM-based methods. This demonstrates that LLMs struggle with extended context reasoning. Although LLM4PFA partially alleviates this issue, the limitation persists.

\textit{(ii) Difficulty with complex cascaded constraints.} When analyzing code containing numerous loops, deeply nested conditional branches, or intricate logical expressions, LLMs often overlook boundary conditions or introduce logical inconsistencies. As a result, they fail to accurately track variable states across multiple reasoning steps. For the cases misclassified by all methods, the average number of conditional statements was 21 higher than the overall dataset average, and 18.9 higher than the subset solvable by LLM4PFA alone. This gap highlights the intrinsic weakness of LLMs in handling multi-step, complex cascaded constraints reasoning.

\textit{(iii) Insufficient semantic understanding.} LLMs also struggle with less common syntactic constructs, pointer manipulations, and user-defined, deeply nested data structures. Their training often lacks sufficient exposure to such patterns, leading to misinterpretation and erroneous reasoning when analyzing these cases.

\vspace{2mm}
\begin{mdframed}[linecolor=gray,roundcorner=12pt,backgroundcolor=gray!15,linewidth=3pt,innerleftmargin=2pt, leftmargin=0cm,rightmargin=0cm,topline=false,bottomline=false,rightline = false]
\textbf{Answer to RQ4 (Case Analysis):}
Although LLMs demonstrate promise in reducing false positives, they still face fundamental limitations in enterprise-scale bug detection. Specifically, LLMs exhibit (1) limited effectiveness in handling long code contexts, (2) difficulty reasoning complex cascaded constraints, and (3) insufficient semantic understanding. 
These challenges remain critical directions for advancing the practical effectiveness of LLMs in future work.
\end{mdframed}

%% file: picture/bug_type.tex
\begin{figure*}[!htbp]
    \centering   
    \subfigure[Accuracy Result using GPT-4o]{
        \includegraphics[width=0.38\linewidth]{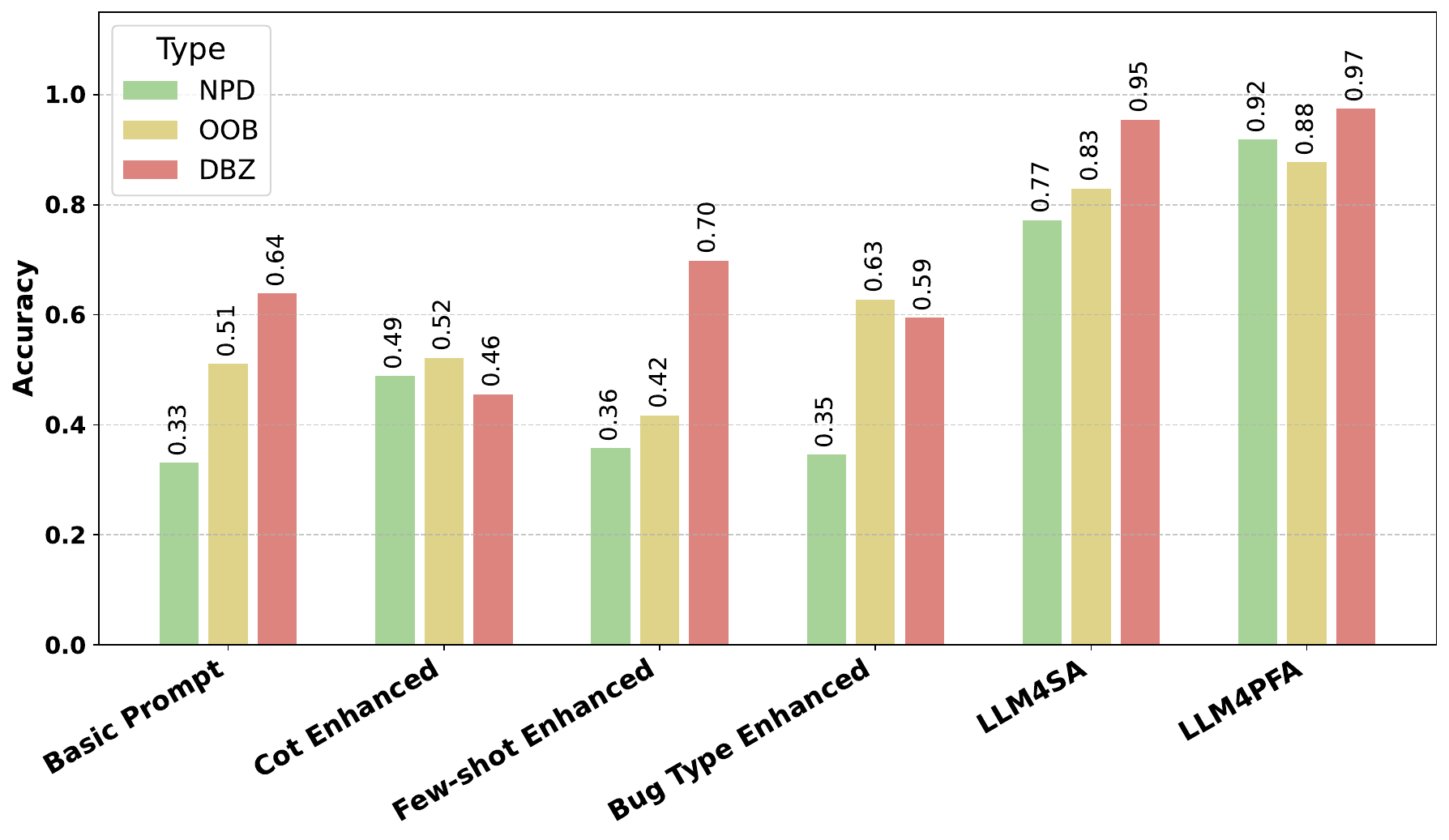}
        \label{fig:gpt4oacc}
    }
    \subfigure[FPR\_Recall Result using GPT-4o]{
        \includegraphics[width=0.38\linewidth]{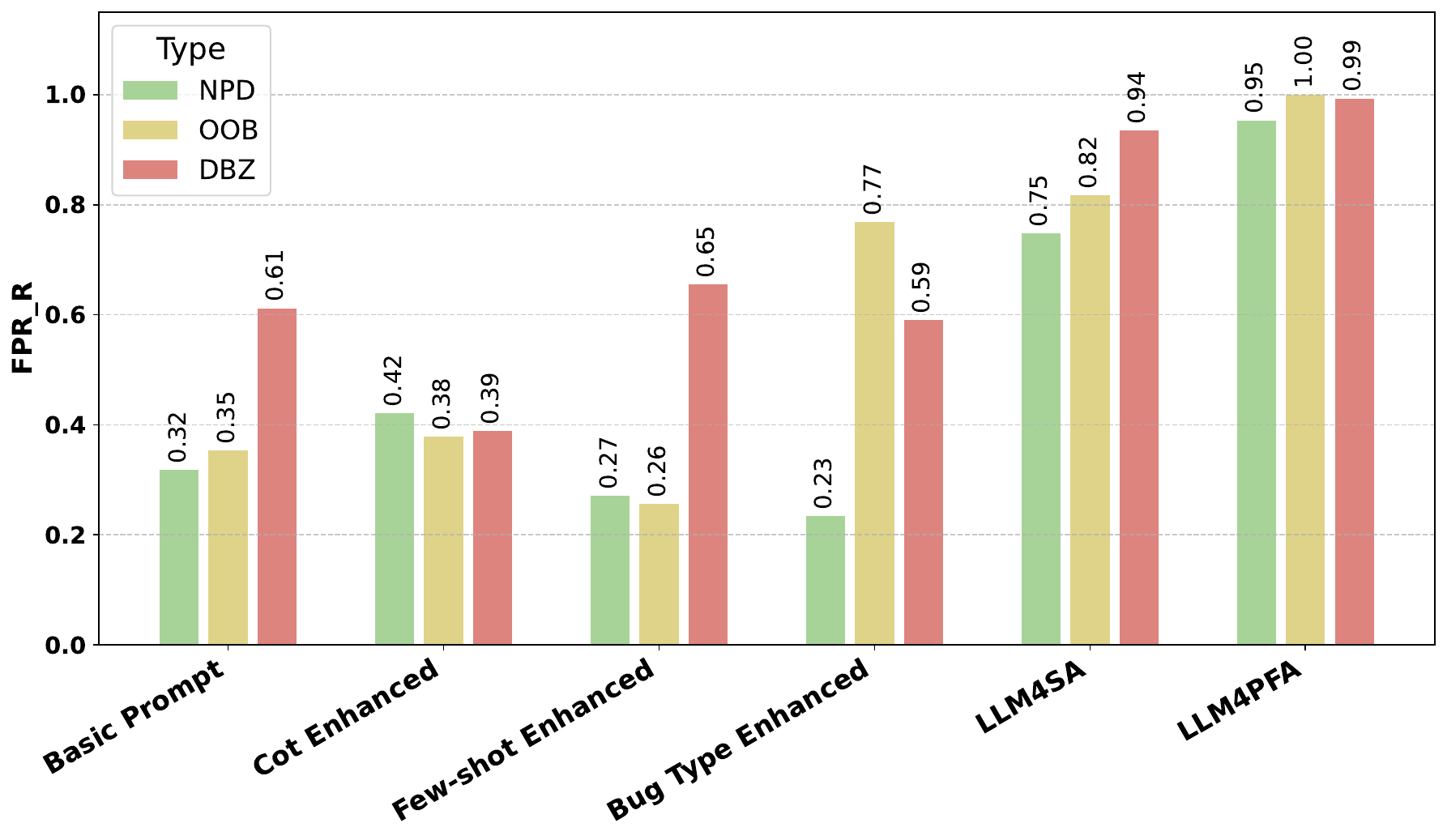}
        \label{fig:gpt4ofpr}
    }   
    \subfigure[Accuracy Result using Qwen-3-Coder]{
        \includegraphics[width=0.38\linewidth]{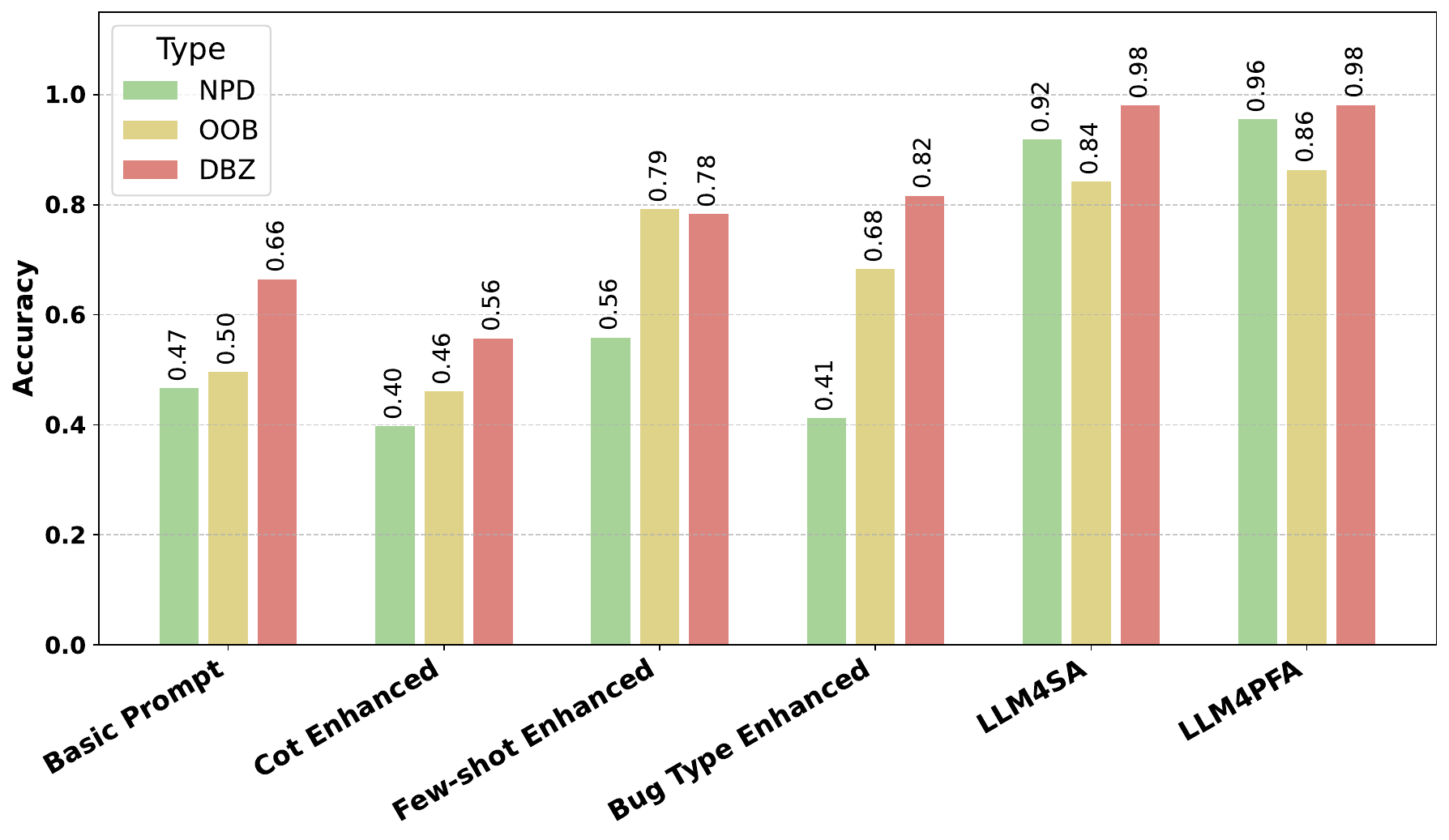}
        \label{fig:qwenacc}
    }
    \subfigure[FPR\_Recall Result using Qwen-3-Coder]{
        \includegraphics[width=0.38\linewidth]{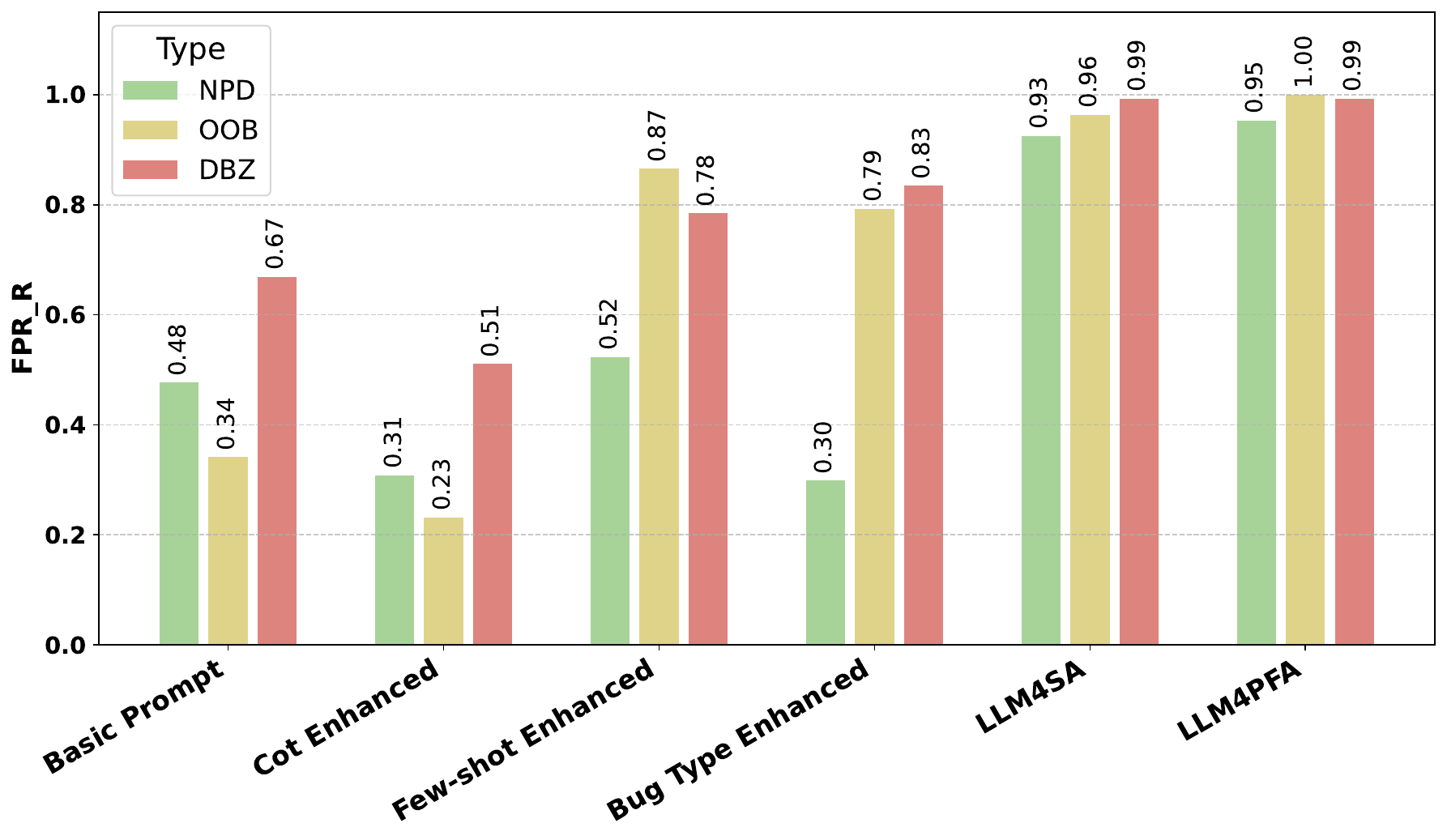}
        \label{fig:qwenfpr}
    }      
    \caption{Comparison of performance for LLM-based methods}
    \label{fig:bugtype_4o}
\end{figure*}

%% file: section/discussion.tex
\section{Discussions}
\label{sec:disccusion}

\subsection{Implications for Practitioners}
\label{sec:Imp_pra}

\subsubsection{How LLMs Help in Practice}
Our results from RQ2-RQ4 demonstrate that several mature LLM-based approaches (i.e., LLM4SA and LLM4PFA) can effectively assist code reviewers in reducing static false positives in real-world enterprise settings. Compared with manually crafted rules, which suffer from poor scalability and limited coverage, and traditional deep learning approaches, which struggle to generalize to proprietary data and often lack interpretability, LLM-based methods offer a more balanced trade-off between effectiveness, interpretability, and efficiency.

\parabf{Effectiveness.} The best-performing LLM-based method LLM4PFA achieved an accuracy of 0.93–0.94 across different state-of-the-art backbone models on our dataset, demonstrating strong potential for false positive reduction. However, recall remains below the enterprise-expected threshold of 90\%, indicating that some degree of manual review is still necessary to guarantee absolute safety. Nevertheless, LLM-based methods can feasibly replace at least one round of manual review in multi-stage validation workflows, reducing overall human effort significantly.

Despite their promise, current LLM-based approaches face several limitations: (1) Long-context reasoning. LLMs still struggle to analyze warnings requiring deep or cross-module context~\cite{longcontext,longcontext2}. The strong performance observed in our dataset partly reflects the limited context sensitivity of the static analyzers used, where most warnings involve only short code fragments. For cases requiring deeper interprocedural reasoning, effectiveness degrades. (2) Complex constraint analysis. LLMs have difficulty accurately simulating program behavior in cases involving cascading constraints~\cite{llm4pfa}. While LLM4PFA alleviates this to some extent, the challenge has not been fully resolved. (3) Bug type coverage. Both prior work and this study focus mainly on memory safety bugs, which exhibit clearer patterns. The performance of LLMs on less common bug types remains an open question. (4) Randomness in reasoning. LLM outputs exhibit randomness. For more stable results, majority voting across multiple runs (e.g., three per case in our evaluation) is recommended.
Moreover, while LLMs show strong potential in filtering false positives, their capacity to discover new bugs remains bounded by the detection capabilities of static analyzers themselves. How to leverage LLMs to help enterprises identify new bugs beyond the coverage of static analyzers remains an open challenge.

\parabf{Interpretability. }
Unlike most traditional deep learning models that output only binary labels, LLM-based methods provide natural language explanations alongside predictions. These explanations significantly support human reviewers, reducing the average time to validate a warning from about 10 minutes to approximately 3 minutes. 
Even when explanations contain inaccuracies, they still aid in quicker issue identification.

\parabf{Efficiency. }As shown in RQ3, all evaluated LLMs perform static false positive reduction at much lower time and acceptable economic costs compared to manual review. Given that model performance is broadly comparable across LLMs, model selection in practice may emphasize efficiency. As such, the open-source Qwen-3-Coder achieves a particularly favorable balance between time and economic cost, while also avoiding the practical challenges of deploying closed-source models in enterprise environments.

\subsection{Implications for Researchers}
\label{sec:Imp_rea}
From a research perspective, our findings highlight several directions for improving the use of LLMs in static bug detection.

\parabf{Knowledge-enhanced LLMs.} RQ2 shows that incorporating domain knowledge significantly improves LLMs reasoning accuracy. Both bug-type descriptions and illustrative examples proved beneficial, with examples yielding greater gains by explicitly exposing recurring bug patterns. In this study, such examples were manually constructed to guide LLM reasoning. Future research could explore automated approaches for acquiring targeted knowledge (e.g., RAG for high-quality historical cases), as well as investigating diverse knowledge sources and modalities for enriching LLMs. These directions may enhance not only effectiveness but also interpretability.

\parabf{Complex contextual and cascaded constraints analysis.} Currently, LLMs demonstrate strong capabilities in analyzing simple code snippets but struggle with scenarios requiring complex context and
cascaded constraints reasoning, which are prevalent in real-world software development. Promising approaches to enhancing LLMs in this direction include decomposing intricate reasoning tasks into smaller, more manageable subproblems where LLMs perform more reliably, or designing hybrid pipelines to support complex contextual analysis.

\parabf{Combining LLMs with Static Analysis.} LLMs and static analysis complement each other effectively. Specifically, LLMs excel at understanding code snippets and generalizing across patterns, so they can enrich static analysis by providing scalable domain knowledge and expanding the coverage of analysis rules. However, LLMs face limitations such as hallucinations and difficulties with long-context reasoning tasks. Static analysis can mitigate these issues by providing reliable program facts and intermediate results, which help reduce hallucinations and assist LLMs in breaking down complex reasoning tasks. The competitive results of LLM4PFA further demonstrate the value of deeply integrating LLMs with static analysis. Thus, combining LLMs with static analysis holds great promise as a future direction for static bug detection.

%% file: section/threats.tex
\section{Threats to Validity}
\label{sec:threats}

\parabf{Construct Validity.}
A potential threat arises from the reliability of manual labeling. To mitigate this, all labels were reviewed through multiple rounds by experienced in-house developers and professional code reviewers, providing strong domain expertise and consistency. Our research team performed further validation, and did not identify any correctness issues in the reviewed labels.

\parabf{Internal Validity.}
Due to the randomness of LLMs, there is an inherent risk to the reliability and consistency of experimental results. To mitigate this, we conducted multiple trials for each LLM-based method and reported the majority outcomes across repeated runs. Moreover, the results of all methods with different LLMs are included in our replication package~\cite{package} to ensure transparency and reproducibility. 
Another threat is the limited dataset scope. Due to enterprise constraints, we collected all available data and applied rigorous filtering to ensure quality. The final dataset includes 433 real cases spanning 10 months, which we believe remains valuable for identifying common false-positive patterns and evaluating the effectiveness of LLMs in false positive reduction.

\parabf{External Validity.}
Our data is collected from Tencent and primarily consists of C/C++ programs, which may limit generalizability to other enterprise environments or languages. Nevertheless, the industrial scale of the dataset still provides meaningful insights for future work. Since the evaluated methods are language-agnostic, they can be extended to other programming languages. In future work, we plan to expand the dataset to include additional business lines and programming languages to further enhance the generalizability of our results.

%% file: section/conclusion.tex
\section{Conclusion}
\label{sec:conclusion}

This work conducts the first comprehensive empirical study of diverse LLM-based false alarm reduction techniques in an industrial context at Tencent, one of the largest IT companies in China. 
Through interviewing developers and analyzing the data, our results highlight the prevalence and bad impact of false positives. Meanwhile, our results show the huge potential of LLMs for reducing false alarms in industrial settings (with best techniques eliminating 94–98\% of false positives). Furthermore, LLM-based techniques are cost-effectiveness, with low time and money costs. Based on findings, we further discuss the practical implications for practitioners and outline promising directions for future research.

%% file: ref.bib
@String{Computing = "Computing" }

@String{Computer = "{IEEE} Computer" }

@String{Springer = "Springer-Verlag" }

@misc{Coverity,
    howpublished = {\url{https://scan.coverity.com}},
    title={Coverity},
    year = {2025},
    key = {coverity}
}

@article{li2023hitchhikers,
      title={The Hitchhiker's Guide to Program Analysis: A Journey with Large Language Models}, 
      author={Haonan Li and Yu Hao and Yizhuo Zhai and Zhiyun Qian},
      year={2023},
      eprint={2308.00245},
      archivePrefix={arXiv},
      primaryClass={cs.SE}
}

@article{llmvulsurvey,
  author       = {Zeyu Gao and
                  Hao Wang and
                  Yuchen Zhou and
                  Wenyu Zhu and
                  Chao Zhang},
  title        = {How Far Have We Gone in Vulnerability Detection Using Large Language
                  Models},
  journal      = {CoRR},
  volume       = {abs/2311.12420},
  year         = {2023},
  url          = {https://doi.org/10.48550/arXiv.2311.12420},
  doi          = {10.48550/ARXIV.2311.12420},
  eprinttype    = {arXiv},
  eprint       = {2311.12420},
  bibsource    = {dblp computer science bibliography, https://dblp.org}
}

@inproceedings{DBLP:conf/icse/YangGMH24,
  author       = {Aidan Z. H. Yang and
                  Claire Le Goues and
                  Ruben Martins and
                  Vincent J. Hellendoorn},
  title        = {Large Language Models for Test-Free Fault Localization},
  booktitle    = {Proceedings of the 46th {IEEE/ACM} International Conference on Software
                  Engineering, {ICSE} 2024, Lisbon, Portugal, April 14-20, 2024},
  pages        = {17:1--17:12},
  publisher    = {{ACM}},
  year         = {2024},
  url          = {https://doi.org/10.1145/3597503.3623342},
  doi          = {10.1145/3597503.3623342},
  bibsource    = {dblp computer science bibliography, https://dblp.org}
}

@article{wei2023chainofthought,
  title={Chain-of-thought prompting elicits reasoning in large language models},
  author={Wei, Jason and Wang, Xuezhi and Schuurmans, Dale and Bosma, Maarten and Xia, Fei and Chi, Ed and Le, Quoc V and Zhou, Denny and others},
  journal={Advances in Neural Information Processing Systems},
  volume={35},
  pages={24824--24837},
  year={2022}
}

@article{zhang2022automatic,
  title={Automatic chain of thought prompting in large language models},
  author={Zhang, Zhuosheng and Zhang, Aston and Li, Mu and Smola, Alex},
  journal={arXiv preprint arXiv:2210.03493},
  year={2022}
}

@article{brown2020language,
  title={Language models are few-shot learners},
  author={Brown, Tom and Mann, Benjamin and Ryder, Nick and Subbiah, Melanie and Kaplan, Jared D and Dhariwal, Prafulla and Neelakantan, Arvind and Shyam, Pranav and Sastry, Girish and Askell, Amanda and others},
  journal={Advances in neural information processing systems},
  volume={33},
  pages={1877--1901},
  year={2020}
}

@misc{sun2023gpt,
      title={When GPT Meets Program Analysis: Towards Intelligent Detection of Smart Contract Logic Vulnerabilities in GPTScan}, 
      author={Yuqiang Sun and Daoyuan Wu and Yue Xue and Han Liu and Haijun Wang and Zhengzi Xu and Xiaofei Xie and Yang Liu},
      year={2023},
      eprint={2308.03314},
      archivePrefix={arXiv},
      primaryClass={cs.CR}
}

@inproceedings{study1fn,
   title={“False negative - that one is going to kill you”: Understanding Industry Perspectives of Static Analysis based Security Testing},
   url={http://dx.doi.org/10.1109/SP54263.2024.00019},
   DOI={10.1109/sp54263.2024.00019},
   booktitle={2024 IEEE Symposium on Security and Privacy (SP)},
   publisher={IEEE},
   author={Ami, Amit Seal and Moran, Kevin and Poshyvanyk, Denys and Nadkarni, Adwait},
   year={2024},
   month=may, pages={3979–3997} }

@inproceedings{buglen,
  title={Towards More Accurate Static Analysis for Taint-Style Bug Detection in Linux Kernel},
  author={Li, Haonan and Zhang, Hang and Pei, Kexin and Qian, Zhiyun},
  booktitle={40th IEEE/ACM International Conference on Automated Software Engineering, ASE 2025},
  year={2025}
}

@inproceedings{study2java,
  title={Comparison and evaluation on static application security testing (sast) tools for java},
  author={Li, Kaixuan and Chen, Sen and Fan, Lingling and Feng, Ruitao and Liu, Han and Liu, Chengwei and Liu, Yang and Chen, Yixiang},
  booktitle={Proceedings of the 31st ACM Joint European Software Engineering Conference and Symposium on the Foundations of Software Engineering},
  pages={921--933},
  year={2023}
}

@inproceedings{study3c,
  title={An empirical study on the effectiveness of static C code analyzers for vulnerability detection},
  author={Lipp, Stephan and Banescu, Sebastian and Pretschner, Alexander},
  booktitle={Proceedings of the 31st ACM SIGSOFT international symposium on software testing and analysis},
  pages={544--555},
  year={2022}
}

@article{DBLP:journals/corr/abs-2503-03586,
  author       = {Alperen Yildiz and
                  Sin G. Teo and
                  Yiling Lou and
                  Yebo Feng and
                  Chong Wang and
                  Dinil Mon Divakaran},
  title        = {Benchmarking LLMs and LLM-based Agents in Practical Vulnerability
                  Detection for Code Repositories},
  journal      = {CoRR},
  volume       = {abs/2503.03586},
  year         = {2025},
  url          = {https://doi.org/10.48550/arXiv.2503.03586},
  doi          = {10.48550/ARXIV.2503.03586},
  eprinttype    = {arXiv},
  eprint       = {2503.03586},
  bibsource    = {dblp computer science bibliography, https://dblp.org}
}

@inproceedings{DBLP:conf/icml/Guo0XS025,
  author       = {Jinyao Guo and
                  Chengpeng Wang and
                  Xiangzhe Xu and
                  Zian Su and
                  Xiangyu Zhang},
  title        = {RepoAudit: An Autonomous LLM-Agent for Repository-Level Code Auditing},
  booktitle    = {Forty-second International Conference on Machine Learning, {ICML}
                  2025, Vancouver, BC, Canada, July 13-19, 2025},
  publisher    = {OpenReview.net},
  year         = {2025},
  url          = {https://openreview.net/forum?id=TXcifVbFpG},
  bibsource    = {dblp computer science bibliography, https://dblp.org}
}

@inproceedings{DBLP:conf/msr/FuT22,
  author       = {Michael Fu and
                  Chakkrit Tantithamthavorn},
  title        = {LineVul: {A} Transformer-based Line-Level Vulnerability Prediction},
  booktitle    = {19th {IEEE/ACM} International Conference on Mining Software Repositories,
                  {MSR} 2022, Pittsburgh, PA, USA, May 23-24, 2022},
  pages        = {608--620},
  publisher    = {{ACM}},
  year         = {2022},
  url          = {https://doi.org/10.1145/3524842.3528452},
  doi          = {10.1145/3524842.3528452},
  bibsource    = {dblp computer science bibliography, https://dblp.org}
}

@inproceedings{DBLP:conf/nips/ZhouLSD019,
  author       = {Yaqin Zhou and
                  Shangqing Liu and
                  Jing Kai Siow and
                  Xiaoning Du and
                  Yang Liu},
  editor       = {Hanna M. Wallach and
                  Hugo Larochelle and
                  Alina Beygelzimer and
                  Florence d'Alch{\'{e}}{-}Buc and
                  Emily B. Fox and
                  Roman Garnett},
  title        = {Devign: Effective Vulnerability Identification by Learning Comprehensive
                  Program Semantics via Graph Neural Networks},
  booktitle    = {Advances in Neural Information Processing Systems 32: Annual Conference
                  on Neural Information Processing Systems 2019, NeurIPS 2019, December
                  8-14, 2019, Vancouver, BC, Canada},
  pages        = {10197--10207},
  year         = {2019},
  url          = {https://proceedings.neurips.cc/paper/2019/hash/49265d2447bc3bbfe9e76306ce40a31f-Abstract.html},
  bibsource    = {dblp computer science bibliography, https://dblp.org}
}

@article{DBLP:journals/corr/abs-2312-05275,
  author       = {Fangzhou Wu and
                  Qingzhao Zhang and
                  Ati Priya Bajaj and
                  Tiffany Bao and
                  Ning Zhang and
                  Ruoyu Wang and
                  Chaowei Xiao},
  title        = {Exploring the Limits of ChatGPT in Software Security Applications},
  journal      = {CoRR},
  volume       = {abs/2312.05275},
  year         = {2023},
  url          = {https://doi.org/10.48550/arXiv.2312.05275},
  doi          = {10.48550/ARXIV.2312.05275},
  eprinttype    = {arXiv},
  eprint       = {2312.05275},
  bibsource    = {dblp computer science bibliography, https://dblp.org}
}

@article{DBLP:journals/corr/abs-2401-17010,
  author       = {Alexey Shestov and
                  Anton Cheshkov and
                  Rodion Levichev and
                  Ravil Mussabayev and
                  Pavel Zadorozhny and
                  Evgeny Maslov and
                  Chibirev Vadim and
                  Egor Bulychev},
  title        = {Finetuning Large Language Models for Vulnerability Detection},
  journal      = {CoRR},
  volume       = {abs/2401.17010},
  year         = {2024},
  url          = {https://doi.org/10.48550/arXiv.2401.17010},
  doi          = {10.48550/ARXIV.2401.17010},
  eprinttype    = {arXiv},
  eprint       = {2401.17010},
  bibsource    = {dblp computer science bibliography, https://dblp.org}
}

@misc{tencent,
  title        = {Tencent},
  howpublished = {\url{https://www.tencent.com}},
  year         = {2025},
  key          = {tencent}
}

@misc{wechat,
  title        = {WeChat},
  howpublished = {\url{https://weixin.qq.com}},
  year         = {2025},
  key          = {wechat}
}

@misc{cloud,
  title        = {Tencent Cloud},
  howpublished = {\url{https://cloud.tencent.com}},
  year         = {2025},
  key          = {cloud}
}

@misc{li2025iris,
      title={IRIS: LLM-Assisted Static Analysis for Detecting Security Vulnerabilities}, 
      author={Ziyang Li and Saikat Dutta and Mayur Naik},
      year={2025},
      eprint={2405.17238},
      archivePrefix={arXiv},
      primaryClass={cs.CR},
      url={https://arxiv.org/abs/2405.17238}, 
}

@misc{juliet,
  title        = {Juliet Test Suite for C/C++},
  howpublished = {\url{https://samate.nist.gov/SARD/test-suites/112}},
  year         = {2025},
  key          = {juliet}
}

@misc{gawk,
  title        = {Gawk},
  howpublished = {\url{https://github.com/gvlx/gawk}},
  year         = {2025},
  key          = {gawk}
}

@misc{cppcheck,
  title        = {Cppcheck},
  howpublished = {\url{http://cppcheck.net/}},
  year         = {2025},
  key          = {cppcheck}
}

@misc{bkcheck,
  title        = {BkCheck},
  howpublished = {\url{https://bk.tencent.com/docs/markdown/ZH/Devops/2.0/UserGuide/Services/Codecc/codecc-intro.md}},
  year         = {2025},
  key          = {bkcheck}
}

@article{llm4sa,
  title={Automatically inspecting thousands of static bug warnings with large language model: How far are we?},
  author={Wen, Cheng and Cai, Yuandao and Zhang, Bin and Su, Jie and Xu, Zhiwu and Liu, Dugang and Qin, Shengchao and Ming, Zhong and Cong, Tian},
  journal={ACM Transactions on Knowledge Discovery from Data},
  volume={18},
  number={7},
  pages={1--34},
  year={2024},
  publisher={ACM New York, NY}
}

@inproceedings{llmdfa,
  title={LLMDFA: Analyzing Dataflow in Code with Large Language Models},
  author={Wang, Chengpeng and Zhang, Wuqi and Su, Zian and Xu, Xiangzhe and Xie, Xiaoheng and Zhang, Xiangyu},
  booktitle={The Thirty-eighth Annual Conference on Neural Information Processing Systems},
  year={2024}}

@misc{gpt4o,
  title        = {GPT-4 Turbo},
  howpublished = {\url{https://platform.openai.com/docs/models/gpt-4-and-gpt-4-turbo}},
  year         = {2024},
  key          = {gpt4o}
}

@misc{claude,
  title        = {Claude 3.5 Sonnet},
  howpublished = {\url{https://www.anthropic.com/news/claude-3-family}},
  year         = {2024},
  key          = {claude}
}

@misc{deepseek,
  title        = {DeepSeek-Coder-V2-Instruct},
  howpublished = {\url{https://huggingface.co/deepseek-ai/DeepSeek-Coder-V2-Instruct}},
  year         = {2024},
  key          = {deepseek}
}

@misc{qwen,
  title        = {Qwen2.5-Coder-32B-Instruct},
  howpublished = {\url{https://huggingface.co/Qwen/Qwen2.5-Coder-32B-Instruct}},
  year         = {2024},
  key          = {qwen}
}

@article{wang2023boosting,
  title={Boosting Static Resource Leak Detection via LLM-based Resource-Oriented Intention Inference},
  author={Wang, Chong and Liu, Jianan and Peng, Xin and Liu, Yang and Lou, Yiling},
  journal={arXiv preprint arXiv:2311.04448},
  year={2023}
}

@misc{infer,
  title        = {Infer},
  howpublished = {\url{https://fbinfer.com/}},
  year         = {2025},
  key          = {infer}
}

@misc{codeql,
  title        = {CodeQL},
  howpublished = {\url{https://codeql.github.com/}},
  year         = {2025},
  key          = {codeql}
}

@inproceedings{cai2023place,
  title={Place your locks well: understanding and detecting lock misuse bugs},
  author={Cai, Yuandao and Yao, Peisen and Ye, Chengfeng and Zhang, Charles},
  booktitle={32nd USENIX Security Symposium (USENIX Security 23)},
  pages={3727--3744},
  year={2023}
}

@inproceedings{cai2021canary,
  title={Canary: practical static detection of inter-thread value-flow bugs},
  author={Cai, Yuandao and Yao, Peisen and Zhang, Charles},
  booktitle={Proceedings of the 42nd ACM SIGPLAN International Conference on Programming Language Design and Implementation},
  pages={1126--1140},
  year={2021}
}

@inproceedings{cai2022peahen,
  title={Peahen: Fast and precise static deadlock detection via context reduction},
  author={Cai, Yuandao and Ye, Chengfeng and Shi, Qingkai and Zhang, Charles},
  booktitle={Proceedings of the 30th ACM Joint European Software Engineering Conference and Symposium on the Foundations of Software Engineering},
  pages={784--796},
  year={2022}
}

@inproceedings{shi2020conquering,
  title={Conquering the extensional scalability problem for value-flow analysis frameworks},
  author={Shi, Qingkai and Wu, Rongxin and Fan, Gang and Zhang, Charles},
  booktitle={Proceedings of the ACM/IEEE 42nd International Conference on Software Engineering},
  pages={812--823},
  year={2020}
}

@inproceedings{shi2018pinpoint,
  title={Pinpoint: Fast and precise sparse value flow analysis for million lines of code},
  author={Shi, Qingkai and Xiao, Xiao and Wu, Rongxin and Zhou, Jinguo and Fan, Gang and Zhang, Charles},
  booktitle={Proceedings of the 39th ACM SIGPLAN Conference on Programming Language Design and Implementation},
  pages={693--706},
  year={2018}
}

@inproceedings{sui2012static,
  title={Static memory leak detection using full-sparse value-flow analysis},
  author={Sui, Yulei and Ye, Ding and Xue, Jingling},
  booktitle={Proceedings of the 2012 International Symposium on Software Testing and Analysis},
  pages={254--264},
  year={2012}
}

@article{vassallo2020developers,
  title={How developers engage with static analysis tools in different contexts},
  author={Vassallo, Carmine and Panichella, Sebastiano and Palomba, Fabio and Proksch, Sebastian and Gall, Harald C and Zaidman, Andy},
  journal={Empirical Software Engineering},
  volume={25},
  pages={1419--1457},
  year={2020},
  publisher={Springer}
}

@inproceedings{yan2018spatio,
  title={Spatio-temporal context reduction: A pointer-analysis-based static approach for detecting use-after-free vulnerabilities},
  author={Yan, Hua and Sui, Yulei and Chen, Shiping and Xue, Jingling},
  booktitle={Proceedings of the 40th International Conference on Software Engineering},
  pages={327--337},
  year={2018}
}

@article{du2024vul,
  title={Vul-RAG: Enhancing LLM-based Vulnerability Detection via Knowledge-level RAG},
  author={Du, Xueying and Zheng, Geng and Wang, Kaixin and Feng, Jiayi and Deng, Wentai and Liu, Mingwei and Chen, Bihuan and Peng, Xin and Ma, Tao and Lou, Yiling},
  journal={arXiv preprint arXiv:2406.11147},
  year={2024}
}

@misc{package,
  title        = {Replication Package},
  howpublished = {\url{https://github.com/LLM4SFPR/static-bug-false-positives-reduction-in-practice}},
  year         = {2025},
  key          = {package}
}

@inproceedings{lee2019classifying,
  title={Classifying false positive static checker alarms in continuous integration using convolutional neural networks},
  author={Lee, Seongmin and Hong, Shin and Yi, Jungbae and Kim, Taeksu and Kim, Chul-Joo and Yoo, Shin},
  booktitle={2019 12th IEEE Conference on Software Testing, Validation and Verification (ICST)},
  pages={391--401},
  year={2019},
  organization={IEEE}
}

@inproceedings{10.1145/3510003.3510153,
author = {Kharkar, Anant and Moghaddam, Roshanak Zilouchian and Jin, Matthew and Liu, Xiaoyu and Shi, Xin and Clement, Colin and Sundaresan, Neel},
title = {Learning to reduce false positives in analytic bug detectors},
year = {2022},
isbn = {9781450392211},
publisher = {Association for Computing Machinery},
address = {New York, NY, USA},
url = {https://doi.org/10.1145/3510003.3510153},
doi = {10.1145/3510003.3510153},
booktitle = {Proceedings of the 44th International Conference on Software Engineering},
pages = {1307–1316},
numpages = {10},
location = {Pittsburgh, Pennsylvania},
series = {ICSE '22}
}

@inproceedings{post2008reducing,
  title={Reducing false positives by combining abstract interpretation and bounded model checking},
  author={Post, Hendrik and Sinz, Carsten and Kaiser, Alexander and Gorges, Thomas},
  booktitle={2008 23rd IEEE/ACM International Conference on Automated Software Engineering},
  pages={188--197},
  year={2008},
  organization={IEEE}
}

@inproceedings{valdiviezo2014method,
  title={A method for scalable and precise bug finding using program analysis and model checking},
  author={Valdiviezo, Manuel and Cifuentes, Cristina and Krishnan, Padmanabhan},
  booktitle={Asian Symposium on Programming Languages and Systems},
  pages={196--215},
  year={2014},
  organization={Springer}
}

@inproceedings{li2013software,
  title={Software vulnerability detection using backward trace analysis and symbolic execution},
  author={Li, Hongzhe and Kim, Taebeom and Bat-Erdene, Munkhbayar and Lee, Heejo},
  booktitle={2013 International Conference on Availability, Reliability and Security},
  pages={446--454},
  year={2013},
  organization={IEEE}
}

@inproceedings{gadelha2019smt,
  title={Smt-based refutation of spurious bug reports in the clang static analyzer},
  author={Gadelha, Mikhail R and Steffinlongo, Enrico and Cordeiro, Lucas C and Fischer, Bernd and Nicole, Denis},
  booktitle={2019 IEEE/ACM 41st International Conference on Software Engineering: Companion Proceedings (ICSE-Companion)},
  pages={11--14},
  year={2019},
  organization={IEEE}
}

@article{llm4pfa,
  title={Minimizing False Positives in Static Bug Detection via LLM-Enhanced Path Feasibility Analysis},
  author={Du, Xueying and Yu, Kai and Wang, Chong and Zou, Yi and Deng, Wentai and Ou, Zuoyu and Peng, Xin and Lou, Yiling},
  journal={arXiv preprint arXiv:2506.10322},
  year={2025}
}

@misc{mao2024effectivelydetectingexplainingvulnerabilities,
      title={Towards Effectively Detecting and Explaining Vulnerabilities Using Large Language Models}, 
      author={Qiheng Mao and Zhenhao Li and Xing Hu and Kui Liu and Xin Xia and Jianling Sun},
      year={2024},
      eprint={2406.09701},
      archivePrefix={arXiv},
      primaryClass={cs.SE},
      url={https://arxiv.org/abs/2406.09701}, 
}

@article{10.1145/3653718,
author = {Wen, Cheng and Cai, Yuandao and Zhang, Bin and Su, Jie and Xu, Zhiwu and Liu, Dugang and Qin, Shengchao and Ming, Zhong and Cong, Tian},
title = {Automatically Inspecting Thousands of Static Bug Warnings with Large Language Model: How Far Are We?},
year = {2024},
issue_date = {August 2024},
publisher = {Association for Computing Machinery},
address = {New York, NY, USA},
volume = {18},
number = {7},
issn = {1556-4681},
url = {https://doi.org/10.1145/3653718},
doi = {10.1145/3653718},
journal = {ACM Trans. Knowl. Discov. Data},
month = {jun},
articleno = {168},
numpages = {34}
}

@misc{li2024llmassistedstaticanalysisdetecting,
      title={LLM-Assisted Static Analysis for Detecting Security Vulnerabilities}, 
      author={Ziyang Li and Saikat Dutta and Mayur Naik},
      year={2024},
      eprint={2405.17238},
      archivePrefix={arXiv},
      primaryClass={cs.CR},
      url={https://arxiv.org/abs/2405.17238}, 
}

@article{primevul,
  title={Vulnerability detection with code language models: How far are we?},
  author={Ding, Yangruibo and Fu, Yanjun and Ibrahim, Omniyyah and Sitawarin, Chawin and Chen, Xinyun and Alomair, Basel and Wagner, David and Ray, Baishakhi and Chen, Yizheng},
  journal={arXiv preprint arXiv:2403.18624},
  year={2024}
}

@article{guo2023mitigating,
  title={Mitigating false positive static analysis warnings: Progress, challenges, and opportunities},
  author={Guo, Zhaoqiang and Tan, Tingting and Liu, Shiran and Liu, Xutong and Lai, Wei and Yang, Yibiao and Li, Yanhui and Chen, Lin and Dong, Wei and Zhou, Yuming},
  journal={IEEE Transactions on Software Engineering},
  volume={49},
  number={12},
  pages={5154--5188},
  year={2023},
  publisher={IEEE}
}

@inproceedings{zhou2024large,
  title={Large language model for vulnerability detection: Emerging results and future directions},
  author={Zhou, Xin and Zhang, Ting and Lo, David},
  booktitle={Proceedings of the 2024 ACM/IEEE 44th International Conference on Software Engineering: New Ideas and Emerging Results},
  pages={47--51},
  year={2024}
}

@inproceedings{10.1145/3639476.3639762,
author = {Zhou, Xin and Zhang, Ting and Lo, David},
title = {Large Language Model for Vulnerability Detection: Emerging Results and Future Directions},
year = {2024},
isbn = {9798400705007},
publisher = {Association for Computing Machinery},
address = {New York, NY, USA},
url = {https://doi.org/10.1145/3639476.3639762},
doi = {10.1145/3639476.3639762},
pages = {47–51},
numpages = {5},
location = {Lisbon, Portugal},
series = {ICSE-NIER'24}
}

@article{Li24Enhancing,
author = {Li, Haonan and Hao, Yu and Zhai, Yizhuo and Qian, Zhiyun},
title = {Enhancing Static Analysis for Practical Bug Detection: An LLM-Integrated Approach},
year = {2024},
issue_date = {April 2024},
publisher = {Association for Computing Machinery},
address = {New York, NY, USA},
volume = {8},
number = {OOPSLA1},
url = {https://doi.org/10.1145/3649828},
doi = {10.1145/3649828},
journal = {Proc. ACM Program. Lang.},
month = {apr},
articleno = {111},
numpages = {26}
}

@article{DBLP:journals/corr/abs-2311-16169,
  author       = {Avishree Khare and
                  Saikat Dutta and
                  Ziyang Li and
                  Alaia Solko{-}Breslin and
                  Rajeev Alur and
                  Mayur Naik},
  title        = {Understanding the Effectiveness of Large Language Models in Detecting
                  Security Vulnerabilities},
  journal      = {CoRR},
  volume       = {abs/2311.16169},
  year         = {2023},
  url          = {https://doi.org/10.48550/arXiv.2311.16169},
  doi          = {10.48550/ARXIV.2311.16169},
  eprinttype    = {arXiv},
  eprint       = {2311.16169},
  bibsource    = {dblp computer science bibliography, https://dblp.org}
}

@inproceedings{regvd,
  title={Regvd: Revisiting graph neural networks for vulnerability detection},
  author={Nguyen, Van-Anh and Nguyen, Dai Quoc and Nguyen, Van and Le, Trung and Tran, Quan Hung and Phung, Dinh},
  booktitle={Proceedings of the ACM/IEEE 44th International Conference on Software Engineering: Companion Proceedings},
  pages={178--182},
  year={2022}
}

@article{longcontext,
  title={Long-context llms struggle with long in-context learning},
  author={Li, Tianle and Zhang, Ge and Do, Quy Duc and Yue, Xiang and Chen, Wenhu},
  journal={arXiv preprint arXiv:2404.02060},
  year={2024}
}

@article{longcontext2,
  title={Make your llm fully utilize the context},
  author={An, Shengnan and Ma, Zexiong and Lin, Zeqi and Zheng, Nanning and Lou, Jian-Guang and Chen, Weizhu},
  journal={Advances in Neural Information Processing Systems},
  volume={37},
  pages={62160--62188},
  year={2024}
}

@article{dl4vd1,
  title={Revisiting the performance of deep learning-based vulnerability detection on realistic datasets},
  author={Chakraborty, Partha and Arumugam, Krishna Kanth and Alfadel, Mahmoud and Nagappan, Meiyappan and McIntosh, Shane},
  journal={IEEE Transactions on Software Engineering},
  volume={50},
  number={8},
  pages={2163--2177},
  year={2024},
  publisher={IEEE}
}

@article{dl4vd2,
  title={Learning-based models for vulnerability detection: An extensive study},
  author={Ni, Chao and Yin, Xin and Shen, Liyu and Wang, Shaohua},
  journal={Empirical Software Engineering},
  volume={31},
  number={1},
  pages={18},
  year={2026},
  publisher={Springer}
}

@phdthesis{dl4vd3,
  title={Understanding and improving deep learning models for vulnerability detection},
  author={Steenhoek, Benjamin Jeremiah},
  year={2024},
  school={Iowa State University}
}

@article{prompt4vd1,
  title={Large language model for vulnerability detection and repair: Literature review and the road ahead},
  author={Zhou, Xin and Cao, Sicong and Sun, Xiaobing and Lo, David},
  journal={ACM Transactions on Software Engineering and Methodology},
  volume={34},
  number={5},
  pages={1--31},
  year={2025},
  publisher={ACM New York, NY}
}

@inproceedings{prompt4vd2,
  title={Exploration on prompting LLM with code-specific information for vulnerability detection},
  author={Liu, Zhihong and Yang, Zezhou and Liao, Qing},
  booktitle={2024 IEEE International Conference on Software Services Engineering (SSE)},
  pages={273--281},
  year={2024},
  organization={IEEE}
}

@article{prompt4vd3,
  title={Llms in software security: A survey of vulnerability detection techniques and insights},
  author={Sheng, Ze and Chen, Zhicheng and Gu, Shuning and Huang, Heqing and Gu, Guofei and Huang, Jeff},
  journal={ACM Computing Surveys},
  volume={58},
  number={5},
  pages={1--35},
  year={2025},
  publisher={ACM New York, NY}
}

@article{study4,
  title={Understanding Industry Perspectives of Static Application Security Testing (SAST) Evaluation},
  author={Li, Yuan and Yao, Peisen and Yu, Kan and Wang, Chengpeng and Ye, Yaoyang and Li, Song and Luo, Meng and Liu, Yepang and Ren, Kui},
  journal={Proceedings of the ACM on Software Engineering},
  volume={2},
  number={FSE},
  pages={3033--3056},
  year={2025},
  publisher={ACM New York, NY, USA}
}
